\documentclass[aps,pre,superscriptaddress,twocolumn,floatfix]{revtex4}

\usepackage[dvips]{graphicx}
\usepackage{amsmath}
\usepackage{amssymb}
\usepackage{natbib}
\usepackage{psfrag}
\usepackage{color}
\usepackage{epsfig}
\usepackage{subfigure}

\pdfoutput=1

\graphicspath{{./}{pdf/}{Figure/}}

\begin{document}

\title{Speed of evolution in large asexual populations with
  diminishing returns }

\author{Maria R. Fumagalli}

\affiliation{Genomic Physics Group, CNRS (UMR 7238) ``Microorganism
Genomics'', 15 Rue de l'\'{E}cole de M\'{e}decine, 75006 Paris, France}

\affiliation{Dipartimento di Fisica,
Universit\`a degli Studi di Milano, Via G. Celoria 16, Milano, Italy}

\affiliation{Dipartimento di Fisica,
Universit\`a degli Studi di Torino, Via P. Giuria 1, Torino, Italy}

\author{Matteo Osella}

\affiliation{Genomic Physics Group, CNRS (UMR 7238) ``Microorganism
Genomics'', 15 Rue de l'\'{E}cole de M\'{e}decine, 75006 Paris, France}

\affiliation{Universit\'e Pierre et Marie Curie, 4 Place Jussieu, 75005  Paris, France.} 

\author{Philippe Thomen}

\affiliation{Universit\'e Pierre et Marie Curie, 4 Place Jussieu, 75005  Paris, France.} 

\affiliation{Laboratoire Pierre Aigrain, Ecole Normale Sup\'erieure, CNRS (UMR 8551), 
 Universit\'e  P. et M. Curie, Universit\'e D. Diderot, 24 rue Lhomond, 75005
Paris, France}

\author{Francois Heslot}

\affiliation{Universit\'e Pierre et Marie Curie, 4 Place Jussieu, 75005 Paris, France.} 

\affiliation{Laboratoire Pierre Aigrain, Ecole Normale Sup\'erieure, CNRS (UMR 8551), 
 Universit\'e  P. et M. Curie, Universit\'e D. Diderot, 24 rue Lhomond, 75005
Paris, France}

\author{Marco Cosentino Lagomarsino}
\email[correspondence to: ]{marco.cosentino-lagomarsino@upmc.fr}

\affiliation{Genomic Physics Group, CNRS (UMR 7238) ``Microorganism
Genomics'', 15 Rue de l'\'{E}cole de M\'{e}decine, 75006 Paris, France}

\affiliation{Dipartimento di Fisica,
Universit\`a degli Studi di Torino, Via P. Giuria 1, Torino, Italy}

\affiliation{Universit\'e Pierre et Marie Curie, 4 Place Jussieu, 75005  Paris, France.}

\begin{abstract}
  The adaptive evolution of large asexual populations is generally
  characterized by competition between clones carrying different
  beneficial mutations.  This interference phenomenon slows down the
  adaptation speed and makes the theoretical description of the
  dynamics more complex with respect to the successional occurrence
  and fixation of beneficial mutations typical of small populations.
  A simplified modeling framework considering multiple beneficial
  mutations with equal and constant fitness advantage captures some of
  the essential features of the actual complex dynamics, and some key
  predictions from this model are verified in laboratory evolution
  experiments.  
  However, in these experiments the relative advantage of a beneficial
  mutation is generally dependent on the genetic background. In
  particular, the general pattern is that, as mutations in different
  loci accumulate, the relative advantage of new mutations decreases,
  trend often referred to as ``diminishing return'' epistasis.  
  In this paper, we propose a phenomenological model that generalizes
  the fixed-advantage framework to include in a simple way this
  feature. To evaluate the quantitative consequences of diminishing
  returns on the evolutionary dynamics, we approach the model
  analytically as well as with direct simulation.  Finally, we show
  how the model parameters can be matched with data from evolutionary
  experiments in order to infer the mean effect of epistasis and
  derive order-of-magnitude estimates of the rate of beneficial
  mutations. Applying this procedure to two experimental data sets
  gives values of the beneficial mutation rate within the range of
  previous measurements.
\end{abstract}

\date{\today}

\maketitle

\section{Introduction}

Although the links between the statistical theory of evolution and
statistical physics can hardly be considered novel~\cite{Park2010}, in
the recent years they are attracting the attention of researchers in a
renewed way.
What changed is the increased availability of genomic and experimental
data, in amounts and levels of precision that (despite there is still
a considerable room for improvement), were difficult to envisage in
the past century.
These innovations are consolidating the biology and genomics of
evolution into a more stable field of application of modeling ideas
from statistical physics.

In particular, contemporary technology allows the realization of
controlled laboratory evolution experiments, which can guide
theoretical investigations and in principle makes the validation and
falsification of phenomenological theories feasible~\cite{Hindre2012}.
These experiments are often performed with large populations of
microorganisms, and allow to explore adaptation under well-defined
sources of natural selection. Moreover, phenotypic characterization
and high-throughput sequencing give a quantitative insight into the
genomic adaptation of these microbial
populations~\cite{Barrick2009,Tenaillon2012a}, with notable
consequences in a wide range of bio-technological and ecological
contexts.
For these large asexual (or rarely mating) populations, a high number
of beneficial mutations emerge in different clones, and cannot be
mixed because of slow or absent recombination. These beneficial
mutations appearing in parallel coexist and compete to drive
adaptation.
This phenomenon of concurrent beneficial mutations (sometimes
generically termed ``clonal interference''), is related to the
Fisher-Muller hypothesis (or Hill-Robertson effect) for the advantage
of recombination~\cite{Felsenstein1974a}.
In general, beneficial mutations also arise with a distribution of
fitness advantage, which is generally believed to be
exponential~\cite{Orr2003}.

Recent models have generally dealt with the competition between
mutations of different strengths and the competition between mutations
that arise on different fitness backgrounds separately.
The first effect, the role of a distribution of fitness effects, is
analyzed by so-called ``clonal interference'' models~\footnote{Note
  that the term has a stricter sense in this case. In the following we
  will reserve the term clonal interference to this stricter meaning
  of competition between mutations of different strength, and talk of
  interference between beneficial mutations in the generic
  case.}~\cite{Gerrish1998,Wilke2004,Park2010}. In these models any
individual is either the wild type or a mutant derived directly from
the wild type. Thus, multiple mutations arising from the extant clones
are neglected.
Conversely, models that explicitly deal with multiple mutations
typically assume that all mutations have the same
effect~\cite{Tsimring1996,Desai2007,Brunet2008,Park2010}.
The latter kind of model has the advantage of being simpler to treat and
accessible analytically, and is characterized by a Gaussian-like
traveling wave for the histogram of log-fitness throughout the
population. In absence of epistasis, this wave moves towards higher
log-fitness with a constant speed and shape. 
New mutations are fixed in the population if they occur in the
high-fitness edge (or ``nose'') of the distribution. Consequently, a
quantitative law relates the width of the log-fitness histogram and
the adaptation speed.
Recent work incorporating both effects~\cite{Good2012} has shown that
if the distribution of fitness advantages given by beneficial
mutations is sufficiently peaked around a single characteristic value,
the evolutionary dynamics can be described by an effective theory with
a single typical selection coefficient and a rescaled beneficial
mutation rate.

While the models described so far have been matched successfully with
the diversity and adaptation speed of short-time laboratory evolution
experiments~\cite{Desai2007a}, there appears to be one important
discrepancy between the models and the behavior of bacteria evolved in
the laboratory for longer times (roughly, $>1000$ generations).
Furthermore, two recent experimental studies~\cite{Khan2011,Chou2011}
have shown a general pattern in the advantage of combined beneficial
mutations occurring in different genes. This combined advantage is
lower than the sum of that of individual mutations. In other words,
when mutations of loci in different genes accumulate, the effective
advantage of each of them is lower.  This was shown by combinatorial
genetics techniques, by constructing all the possible configurations
of a small set of mutations, and evaluating their advantage through
competition experiments.
This trend, referred to as ``diminishing returns'' epistasis, had been
previously suggested theoretically on the basis of the general pattern
of adaptation observed in long-term microbial
experiments~\cite{Kryazhimskiy2009}, using a modeling framework that
neglected concurrent or multiple mutations.
Another study predicts the same principle on the basis of a simple 
fitness landscape model combined with the distribution of single
mutation effects measured experimentally~\cite{Martin2007}.
The actual pattern in the fitness associated to the same mutation in
different backgrounds observed by the two studies is complex, as, on
top of the diminishing return effect, the advantage appears to depend
on the mutation identity.
Based on their data, Chou and coworkers~\cite{Chou2011} defined an
expression for the fitness made of two additive components.
They consider the growth rate as the sum of two ``constituent
phenotypes'': metabolic rate and protein expression burden. When
mutations are combined, the metabolic and expression contributions to
the fitness combine multiplicatively in an independent way.
They show that this model generates an accurate prediction for the
fitness of all their multi-allele strains.
Even more recent systematic experiments~\cite{Tenaillon2012a} are
unveiling a complex scenario where different mechanisms coexist for
the interactions of mutations between and within functional
``blocks'', which can span multiple genes along the genome.
One interesting modeling approach that was developed
recently~\cite{Schiffels2011} incorporates genetically linked multiple
mutations explicitly, and describes interference interactions between
multiple beneficial and deleterious mutations. This description has
enough degrees of freedom to possibly account for many of the
phenomena observed experimentally.
However, the full experimental complexity is difficult to incorporate
in a treatable model, and experimental data on linked mutations
and interference between them are not easy to be obtained. 
A simplified descriptions in the spirit of
the multiple mutations model can be useful in order to model
evolving populations using a minimal quantity of information
on mutations and fitness advantage.

	Here, we take a simplified approach to investigate the diversity and
speed of adaptation in presence of diminishing returns.
We define a framework that can account for multiple mutations, and
incorporates the effect of diminishing return epistasis in a
simplified way. 
Namely, the fitness of a mutation depends only on its order of
appearance in a clone, and decreases with it.  This generalizes the
standard multiple mutation model, which we recover in the case in
which the fitness decrease with mutation index is zero.  
We preserve the model assumption that evolution is driven by
beneficial mutations which appear with constant rate (see the
Discussion for an evaluation of these assumptions in light of the
results).
We study the mean-field behavior of this model using standard
techniques~\cite{Park2010}, and derive a relation that is analogous to
Fisher's theorem. Moreover, we discuss how the model assumptions cease
to be valid when the rate of fixed beneficial mutations drops to the
level where deleterious mutations should be accounted for.
We also explore the behavior at finite population sizes, and obtain
simple analytical quantitative estimates of the speed of adaptation
that generalize that of the standard multiple mutation model.
Finally, we show how the model parameters can be matched with data
from two different long-time laboratory evolution experiments and how,
assuming the model, one can derive order-of-magnitude estimates of the
beneficial mutation rate and the mean effect of the diminishing
returns.

\section{Basic features of the model}

\subsection{Model definition}

\begin{figure}[htb]
  \centering
  \includegraphics[width=0.40\textwidth]{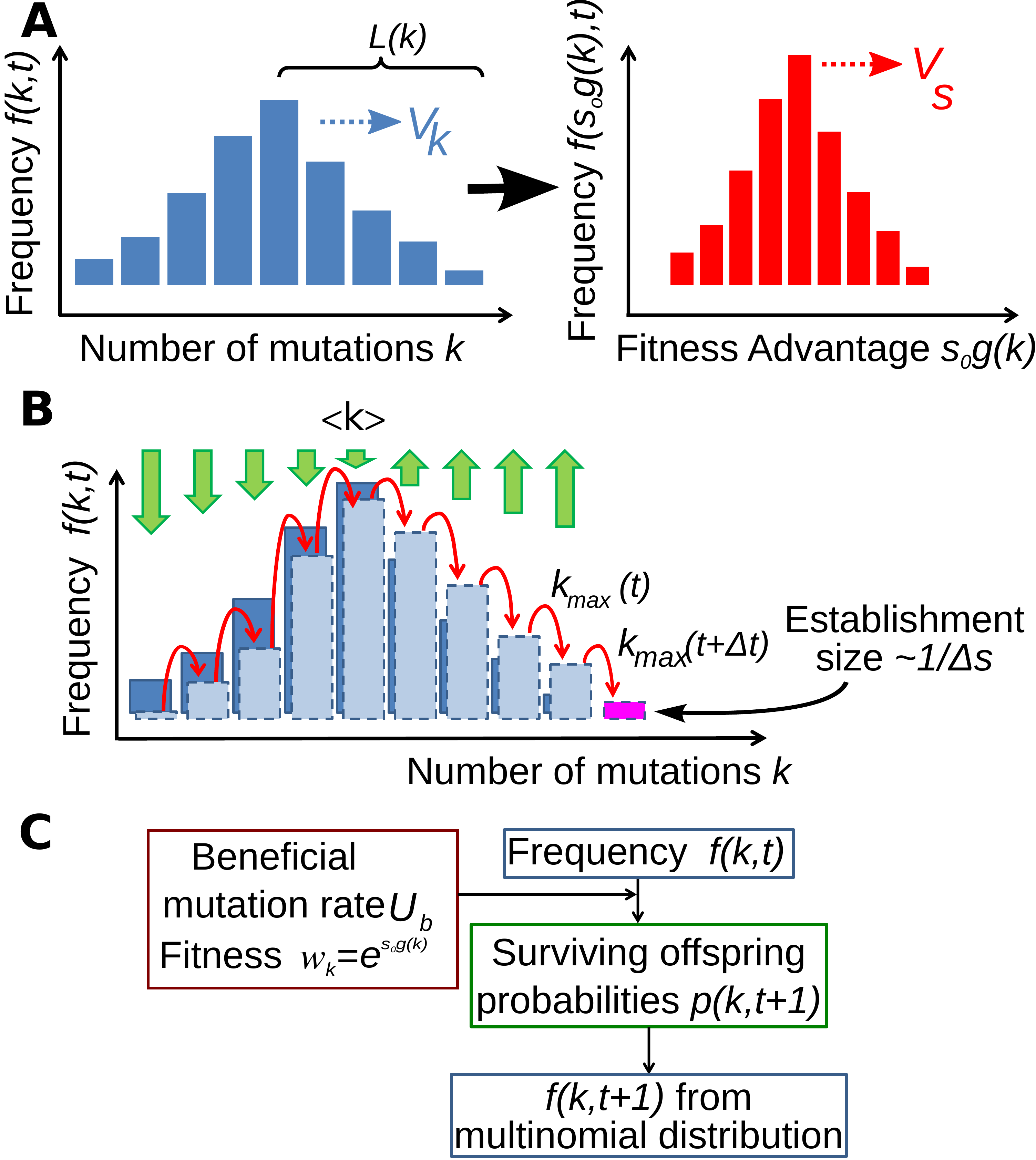}
  \caption{ (Color online) \textbf{Basic features of the model.} (A)
    Because of competition between beneficial mutations, the
    population is divided into sub-populations with different
    frequencies (left panel), defined by the number of mutations $k$.
    $L_k$ is the difference, in number of mutations, between
    the maximum number of mutations found in a clone $k_\mathrm{max}$
    and the mean. This induces a distribution for 
    the log-fitness $s_0 g(k)$ (right panel). Both distributions
    travel in time, driven by established beneficial mutations, with
    instantaneous velocities $v_k$ and $v_s$.
    (B) The dynamics is driven by beneficial mutations and the rate of
    establishment of new classes.  The dashed rectangles represent the
    class frequency histogram at a subsequent time.  
    (C) Sketch of the algorithm used in the simulations (from
    ref.~\cite{Park2010}, see text). The instantaneous frequencies of
    mutation classes define, through Eq.~\ref{eq:algorithm}
    (which incorporates selection and mutation) the probability $p(k,t+1)$
    that an individual with $k$ mutations is found at the subsequent
    generation.  The frequencies at the subsequent generations are
    then sampled from a multinomial distribution with parameters
    $p(k,t+1)$. This procedure allows to access large population sizes.
\label{fig1}
}
\end{figure}

We model an asexual population of $N$ haploid individuals, or
sequences, in which each individual of type $i$ produces a random
number of offspring with average equal to its fitness $w_i$.
Inheritance is introduced by assigning the fitness of their parents to
the offspring.  The evolutionary dynamics of the population is based
on the well-known Wright-Fisher model~\cite{Wright1931,Fisher1930}. We
used the following reproduction scheme~\cite{ParkKrug}.
Each individual at generation $t+1$ is chosen as the offspring of an
individual in class $i$ present at generation $t$ with probability
$\chi_i/ N $, where $\chi_i = w_i/\langle w\rangle $ is the relative
fitness of class $i$. 
This prescription for the evolutionary dynamics 
assumes non-overlapping generations since at each generation there is
a complete replacement of parents with progeny. The population size
$N$ is kept constant, and there is no recombination move.  The
presence of fitness differences in the population leads to natural
selection since classes of individuals with higher fitness will
generate increasingly larger fractions of the population as the
dynamics proceeds, while classes with low fitness will progressively
disappear.

Each offspring has a constant probability per generation $U_b$ (the
``beneficial mutation rate'') of acquiring a beneficial
mutation. Beneficial mutations increase the parental fitness $w_i$ to
a higher value $w'_i$ following the relation $ w'_i=w_i(1+s)\simeq w_i
e^s$, where the positive parameter $s$ is the ``selection
coefficient'' whose typical values in laboratory evolution experiments
with bacteria are in the range
$s\simeq0.001-0.005$~\cite{Hegreness2006,Perfeito2007}.  
We assume that each new beneficial mutations hits a new site on the
genome.
While the fitness of new mutations is a complex
issue~\cite{Eyre-Walker2007}, in presence of abundant beneficial
mutations, deleterious mutations (negative effect on fitness, $s<0$)
do not typically contribute to the adaptation of large populations and
are customarily neglected~\cite{Park2010,Desai2007,Brunet2008}.

The conditions for the emergence of the interference phenomenon
between mutations for a large population can be understood with simple
scaling arguments as a competition between processes occurring at
different time scales~\cite{Park2010}.  When a new mutant with
advantage $s$ arises, the probability that its lineage grows
sufficiently in size to overcome genetic drift (stochastic
fluctuations in the reproductive process) and start to
deterministically expand in the population (i.e. it is
``established'') is $\pi(s) \simeq c s$, where c is a constant factor
that depends in the specific model used~\cite{Haldane1927}. For the
algorithm used here, $\pi(s) \sim 2s$ (see Appendix,
Fig.~\ref{fig_PI})~\footnote{ see also ref.~\cite{Park2010} for a
  self-contained motivation of these formulas.}.
Furthermore, since the extinction probability is $1-\pi(s)$, one can
also estimate the population size conditioned to certainty of survival
of the lineage as $1/\pi \simeq 1/2s$.
In absence of additional beneficial mutations, once a lineage is
established (i.e. has survived genetic drift and has roughly size
$1/2s$) it will take over the population (go to ``fixation'')
logistically at initial rate $s$. 
Consequently, the scale of the
fixation time can be estimated by imposing that  $\frac{1}{2s}\exp(s
\tau_{\mathrm{fix}}) \simeq N$, giving a characteristic time
$\tau_{\mathrm{fix}}\simeq \frac{ln(2Ns)}{s}$~\cite{Desai2007}.
On the other hand, the time scale for appearance and establishment of
a new beneficial mutation is $\tau_{\mathrm{est}}\simeq \frac{1}{N U_b
  \pi(s)}$. Therefore, when $ N U_b\ll \frac{1}{2ln(2Ns)}$
($\tau_{\mathrm{fix}} \ll \tau_{\mathrm{est}}$) a beneficial mutation
can fix before any new mutation can establish, making the evolutionary
dynamics driven by successive sweeps of new lineages arising on an
essentially clonal population. This regime is called ``selective
sweeps'' or ``periodic selection''.
Instead, a sufficiently large population with high beneficial mutation
rate evolves in the opposite regime, in which multiple mutations can
establish before fixation of any of them and interfere with each
other.  This is the regime considered here, which is believed to be
relevant for laboratory evolution experiments with microorganisms.

As mentioned in the introduction, the standard approach of multiple
mutation models is to consider constant advantage $s=s_0$ for all
mutations.
This entails that, in absence of epistasis, the fitness $s_0 k$ can be
assigned to individuals (sequences) with $k$ mutations, which
(roughly) propagate with rate $1 + s_0 (k -\langle k
\rangle)$~\cite{Brunet2008}. $\langle k\rangle$ is the average number
of accumulated mutations in a single realization of the process,
i.e. the average over the distribution of $k$ shown in
Fig.~\ref{fig1}.
For a range of parameter values, this constant advantage framework can
be seen as an effective theory for a model with a distribution of
fitness for beneficial mutations~\cite{Good2012}.

Following the spirit of this modeling framework, we build a minimal
phenomenological model including diminishing return epistasis in
presence of competition between beneficial mutations. The model
assumes that successive mutations do not lead to the same fitness
gain, but the fitness gain is dependent on the number of mutations
already occurred in an individual~\cite{Kryazhimskiy2009}.  A simple
way to implement this feature is to extend the standard model for
multiple mutations and consider selection coefficients dependent on
the number of mutations, i.e.  $s=s_0 g'(k)$, where $g'(k)$ is a
decreasing function of $k$.  As in the standard multiple-mutation
model~\cite{Desai2007,Brunet2008}, the population can be divided into
classes of individuals with the same number of mutations that are in
one-to-one correspondence to fitness advantage classes, as represented
in Fig.~\ref{fig1}A.
An individual with $k$ beneficial mutations has fitness
\begin{equation}
  w_k=e^{\left( \sum_{k'=0}^k s_0 g'(k') \right)} = e^{s_0 g(k)} \ .
\label{fitness_def}
\end{equation}
Since the rules defining the dynamics contain the \emph{relative}
fitness $\chi_k$, the model is unaffected by multiplication of all the
$w_k$ by a common factor. Therefore, the fitness value $w_0=1$ can be
arbitrarily assigned to the genotype with no mutation (``ancestral''
or ``wild-type'').

We considered typical values of the parameter $s_0$ between $10^{-3}$
and $10^{-1}$~\cite{Perfeito2007,Park2010}. For the parameter $U_b$,
we explored the range
$10^{-10}-10^{-3}$~\cite{Perfeito2007,Hegreness2006}, and we
considered population sizes between $10^6$ and
$10^{10}$~\cite{Hindre2012}.
These ranges, which can be considered plausible with respect to what
is known empirically, impose a hierarchy of scales in the effective
parameters. For example, typically $U_b \ll s_0$ and $s_0 N$ is
large. As for $U_bN$, we are interested in exploring sufficiently
large values to ensure the regime of concurrent mutations.
However, since the advantage given by a beneficial mutation $s_0
g'(k)$ decreases during the time-evolution of the model, this
hierarchy can in part change over time. When the advantage becomes too
small, i.e. for large times, some of the assumptions of the model
become unrealistic. Namely, for sufficiently large times, deleterious
mutations cannot be neglected, and the establishment size $1/\pi(s)$
can eventually become comparable to $N$.  The inequality $ N U_b\gg
\frac{1}{2ln(2Ns)}$, which ensures concurrent mutations, can also
break down for very large times, when $s$ decreases.
This is not a problem - as any experiment spans a finite time.
However, in contrast to the standard multiple mutation model, which
describes a stationary state where the regime is entirely defined by
the parameters, in this model the time-range of validity of the
assumptions has to be kept in mind. We will comment on these issues in
the following sections, when reviewing the results.

In order to fully define the model, one has to choose a specific functional
form for the function $g'(k)$, describing the strength of the negative
epistasis between mutations.  
In general, every function $g'(k)$ leading to a sum
$g(k)=\sum_{k'=0}^k s_0 g'(k') $ that is sub-linear in $k$ defines a
model with diminishing return.  A simple example is given by the
choice of a fitness gain that depends on the number of the extant
mutations $k$ as a power law.  In this case, $g'(k) =\alpha k^{\alpha
  -1}$ with $\alpha \le 1$, where the epistasis grows in strength with
decreasing $\alpha$ from the non-epistatic case $\alpha=1$.
In the case $0< \alpha <1$, the fitness of an individual with $k$
beneficial mutations (see Appendix~\ref{appendix:modelli} for the
derivation) has the form

\begin{equation}
w_k=e^{s_0 k^\alpha} \ .
\label{eq:modello-plaw}
\end{equation}

Appendix~\ref{appendix:modelli} describes two additional examples of
the diminishing returns function $g'(k)$ that we have considered,
leading to logarithmic (i.e. the case $ \alpha = 0 $) or geometric
increase of the log-fitness. The comparison between the three variants
will be useful in the second part of this work, when the model
parameters are matched with experimental data.

\subsection{Background on the multiple mutation model with no
  epistasis}

For $\alpha = 1$, the power law return model reduces to the particular
case of absence of epistasis (i.e. $g'(k)=1$, hence $w_k=e^{k
  s_0}$). This case coincides with the standard multiple mutation
model for the evolutionary dynamics of a large asexual
population~\cite{Rouzine2003,Desai2007}, in which every mutation
carries a constant fitness advantage $s_0$.
The phenomenology of the multiple mutation model with constant
advantage has been explored in detail
theoretically~\cite{Desai2007,Brunet2008} and some of the resulting
predictions have been experimentally tested~\cite{Desai2007a}.

  As mentioned in the Introduction, one of the main results of these
studies is that a steady state exists, where the population is
organized in fitness classes (and the corresponding mutation classes,
as illustrated in Fig.~\ref{fig1}A), forming an approximately Gaussian
traveling wave encompassing a constant number $2L$ of
mutation-classes and propagating to higher average fitness with
constant velocity $v_s$. Technically, this wave has all the
characteristics of a soliton~\cite{Rouzine2003,Park2010}.
Importantly, the wave is driven by the mutations that are fixed at the
edge of the distribution of the advantage. Fig.~\ref{fig1}B depicts
schematically how the fitness and mutation class waves propagate.
Mutation class $k$ is fed by beneficial mutations from the previous
class, $k-1$. The fittest class $k_{max}$ produces a new class of
mutants that will be subjected to genetic drift until its size reaches
the establishment size that depends on its relative
advantage. Assuming new class $k_{max}+1$ senses a constant background
from the class with mean fitness, the advantage can be estimated as
$s_0 L$, where $L$ indicates its distance from $\langle k \rangle$
(see Fig.~\ref{fig1}).
Under this argument, the establishment size is $1/2\Delta s\sim 1/2s_0
L$.  During the time in which a new fittest class establishes, the
peak of the distribution will advance and the less-fit clones die out,
thus keeping constant the distribution width (i.e. constant $L$).

The speed of adaptation can be estimated by a scaling
argument~\cite{Desai2007,Brunet2008,Park2010} imposing that the time
$\tau$ during which a new class establishes (and the class histogram
moves by one bin) has to correspond to the larger-scale movement of
the histogram by deterministic growth.

The first part of the argument estimates $\tau$ from the condition
\begin{equation}
  1 \simeq \int_0^{\tau}\mathrm{d}t 
           \frac{U_b}{2s_0L} e^{s_0 (L-1) t} \pi(s_0L)     \ ,
\label{eq:DF1}
\end{equation}
which imposes that order one mutations are produced (and establish
with probability $\pi$) from the foremost bin while it grows
exponentially according to its advantage.

 The foremost bin is treated as
it were growing in a constant background with mean fitness (as above)
and $1/2s_0L$ is taken as the size of the foremost bin at its
birth. Integration of the above equation yields the expression 
\begin{equation}
  \tau \simeq \frac{1}{s_0 L} \log \frac{s_0 L}{U_b} \ ,
\label{eq:DF1_tau}
\end{equation}
which is valid for sufficiently large $L$ and advantage of the fittest
class, since the contribution of the integration boundary $t=0$ has
been neglected.

The second part of the argument can be formulated as a normalization
condition of the fitness class histogram, joint with the condition
that each bin grows exponentially according to its advantage, 
\begin{eqnarray}
  \frac{N}{2} \simeq  \frac{1}{2s_0L} \left(   1 + e^{s_0(L-1)\tau} +
    e^{s_0((L-1)+(L-2))\tau} + \ldots  \right) \nonumber \\ =  
    \frac{1}{2s_0L} \left(1+ \sum_{j=1}^{L-1} e^{s_0\tau \sum_{i=1}^j
        (L-i)}  \right);\:
\label{eq:DF2}
\end{eqnarray}

note that this condition neglects the contribution of beneficial
mutations from the neighboring bins. Approximating the above formula
as an integral computed with saddle-point (i.e. assuming that the
largest term in the sum, the class with mean fitness,
dominates~\cite{Desai2007,Brunet2008}) yields the expression
\begin{equation}
  \label{eq:DF2_tau}
  \log(s_0 L N) \approx \frac{s_0 \tau}{2} L^2.
\end{equation}

Equating the two expressions Eq.~(\ref{eq:DF1_tau}) (or some variants
that can be derived more rigorously) and (\ref{eq:DF2_tau}) for $L$ or
$\tau$ generates implicit estimates for the speed of the mutation
classes histogram, $v_k = 1/\tau$, which in general work very well,
despite of the approximations taken.
In particular, neglecting the addends $\log L$ in
Eq.~(\ref{eq:DF1_tau}) and~(\ref{eq:DF2_tau}), i.e. up to logarithmic
corrections in $L$, one obtains the closed
formula~\cite{Desai2007,Brunet2008}
\begin{equation}
  \label{eq:DFclosed}
  \tau \approx \frac{1}{v_k} \approx \frac{(\log s_0/U_b)^2}{2 s_0 \log
    (Ns_0)} \ .
\end{equation}
Finally, $v_s$ can be estimated by $s_0/\tau$. 

In the general case of diminishing return epistasis, the variables of
interest become a function of the genetic background. Specifically,
the speed of adaptation in log-fitness space $v_s$ and the speed $v_k$
of the mutation classes, are two non-trivially distinct quantities,
and the same holds for the width of the mutation and the advantage
histograms ($L_k$ and $L_s$) relative to the population
(Fig.~\ref{fig1}).

\subsection{Simulation algorithm and effective parameters}

The model was simulated with the algorithm of Park and Krug, as
described in refs~\cite{ParkKrug,Park2010}. The simulation scheme is
sketched in Fig.~\ref{fig1}C. In a typical initial configuration, all
clones have $k=0$ mutations. At each subsequent time step the progeny
of the individuals of all fitness classes are sampled from a
multinomial distribution of parameters $\{p(k,t+1)\}_{\{k \in
  [k_{\mathrm{min}},k_\mathrm{max}] \} }$.  The parameters $p(k,t+1)$
take into account the relative fitness $\chi_k$ computed at time $t$
and the contribution of beneficial mutations. Specifically 
\begin{equation}
p(k,t+1) =(1-U_b)f(k, t) \chi_k(t) \:+ \: U_bf(k-1,t) \chi_{k-1}(t)\ ,
  \label{eq:algorithm}
\end{equation}
\\
(see also Eq.~\eqref{freq_meanfield}).  Together, these definitions
are equivalent to a Wright-Fisher model with separate selection and
mutation steps. In particular, the microscopic step described above
reduces to the standard Wright-Fisher model for genetic drift when
$U_b=0$.  The multinomial random numbers are generated by iteratively
drawing binomial random numbers with parameters $q(k,t+1) = p(k,t+1) /
\sum_k p(k,t+1) $ starting from $k_{\mathrm{max}}$.

The model defined above is invariant by suitable rescaling of time,
provided the other model parameters are also rescaled correctly.  This
feature is useful in comparison with experiments (see following) in
order to understand the correspondence between time steps in the model
and a generation in the experiment. It can also be useful in
increasing the efficiency of simulations.

Let us suppose one desires to map a reference empirical time into the
time steps of the model. This requires the mapping
$t_{\mathrm{model}}=r t_{\mathrm{emp}}$, in the expression describing
the growth of a sub-colony from a clone in terms of the experimental
generation time.  The coefficient $r$ is the number of empirical
generations corresponding to a time step, and represents the
simulation time scale.  
A simple choice of time scale is the case $r=1$, implying a one-to-one
correspondence between empirical generations and time steps in the
model.
Since an established clone with advantage $s$
grows as $e^{s t}$, the advantage function is proportional to the time
scale as follows, $s_{\mathrm{model}}= r \:s_{\mathrm{emp}}$.  This
means that the value of the fitness in the model depends exponentially
on $r$.  
Similarly, since the beneficial mutation rate is defined as the number
of expected mutations per genome per generation, the map between time
steps and empirical generations implies $U_{b,\mathrm{emp}}=
U_{b,\mathrm{model}}/r$.
Finally, the correct rescaled population size is
$N_{\mathrm{model}}=N_{\mathrm{emp}}/r$. In practice $N$ can be rather
large in a typical experiment (e.g. $\approx 10^{9}$), so that the
rescaling of population size does not affect much the dynamics
provided $r$ is not too large.
The rescaling of the parameters described above can easily be
rationalized keeping in mind that the basic time scales of the model
are set by the products $sN$ and $N U_b$.
In order to verify that the invariance discussed above is valid for
our model, where the advantage $s$ varies with the number of mutations
as $s_0 g(k)$, we ran some simulations choosing the model parameters
($N$, $s_0$, $U_b$) using different maps between the time units (see
Appendix, Fig.~\ref{fig_R}). The results indicate that for any
practical purpose, the invariance is effective.

\section{Results}

\subsection{Adaptation slows down until it reaches an effective
  arrest.}

\begin{figure}[htbp]
  \includegraphics[width=0.49\textwidth]{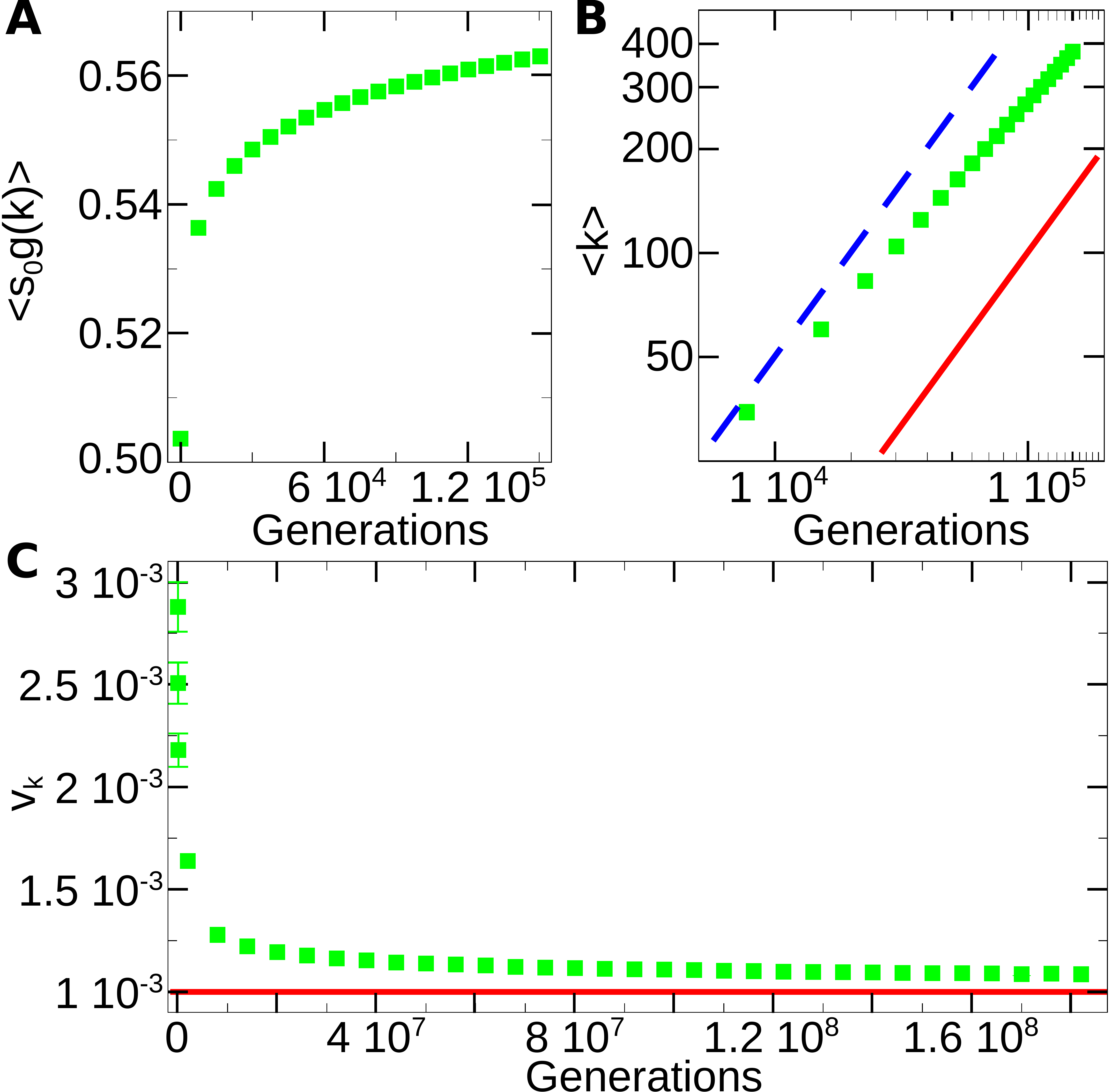}
  \caption{ (Color online) \textbf{The mean number of mutations and the
      mean advantage grow sub-linearly with time.}  The top panels
    show the increase in time of $\langle s_0g(k) \rangle$ (A) and of
    $\langle k \rangle$ (B) obtained by direct simulation of the
    diminishing return model. The plot in panel B is in log-log scale,
    and the data are compared with a reference line (dashed blue line,
    with slope $5\cdot10^{-3}$) to highlight the sublinear growth of
    $\langle k\rangle$. The continuous red line shows the the
    asymptotic long-time linear behavior with slope corresponding to
    $U_b$.  (C) Long-time behavior of the mean speed of fixed
    mutations $v_k$ (green symbols), averaged over different
    realizations.  For long times, this quantity decreases (as a power
    law) towards the limit value $v_k=U_b$, where the assumptions of
    the model break down and deleterious mutations need to be
    accounted for~\cite{Trav_wave}. This limit also corresponds to the
    limit value of $v_k$ obtained by a mean-field estimate (see text).
    Simulations are carried out using the parameters $N=5\cdot10^7$,
    $s_0=0.5$, $\alpha=0.02$, $U_b=1 \cdot 10^{-3}$.  Averages are
    computed over $100$ realizations (these averages are implied in
    the notations for the y-axis labels).}
  \label{fig_KWV}
\end{figure}

Direct simulation of the model (Fig.~\ref{fig_KWV}A), shows that the
mean advantage $\langle s_0g(k) \rangle$ grows sub-linearly with time,
as expected from the diminishing returns pattern of the advantage.
The average number of mutations $\langle k \rangle$ also grows
sub-linearly with time, for intermediate to long times (Fig.~\ref{fig_KWV}B).
This trend is independent from $\alpha$ (or from the specific model of
the of decreasing $g'(k)$ 
and is due to the fact that decreasing advantage and the consequent
rise of the establishment threshold for clones together slow down
adaptation. The time derivatives of $\langle s_0 g(k) \rangle $ and
$\langle k \rangle$ estimate the adaptation speed $v_s$ and the
mutation-accumulation speed $v_k$ of a typical realization.
Figure~\ref{fig_KWV}C shows an average of $v_k$ over one hundred
realizations, plotted as a function of time.
The simulations indicate that $v_k$ relaxes to a plateau which is
close to the beneficial mutation rate $U_b$. Equivalently, for long
times, the mean number of fixed mutations shows a linear behavior in
time with a rate close to $U_b$ (red line in Fig.~\ref{fig_KWV}B).  In
the same long-time limit, the advantage of a mutation $s=s_0 g'(k)$
drops asymptotically to zero.

The long-time regime where $\langle k \rangle$ follows a linear trend
is outside of the limit of validity of the model, and has to be
regarded as unphysical, since when $v_k = U_b$, deleterious mutations
cannot be neglected~\cite{Trav_wave}. 
Thus, for any finite $N$ the asymptotic trend of $\langle k \rangle$
has to be interpreted as an effective signature of an evolutionary
arrest for both $v_k$ and $v_s$, where beneficial mutations should be
in equilibrium with deleterious one, which could possibly be captured
by a variant of the model including deleterious
mutations~\cite{Trav_wave}. Note also that realistically the
beneficial mutation rate itself could decrease in the later stages of
evolution~\cite{Park2010,Park2008}.

Despite the fact that the assumptions of the model break down
asymptotically, the long-time phenomenology appears to be
theoretically consistent. Indeed, the long-time regime can be easily
rationalized as a transition to an ``effectively neutral'' evolution
regime.  In fact, the selection coefficient $s_0 g'(k)$ becomes
irrelevant at long times, so that the fixation dynamics is driven
solely by genetic drift.  Given that deleterious mutations are not
accounted for, and the benefit carried by the allowed mutations
becomes negligible, in this regime $U_b$ effectively becomes a neutral
mutation rate.
The probability of fixation of an essentially neutral mutation is
$\sim 1/N$ while the rate of appearance of new mutations is $U_b
N$. Therefore, the pace at which new mutations are accumulated is
approximately $v_k\sim U_b$.  As we will see, this asymptotic velocity
$U_b$ is also recovered from a mean-field estimate of $v_k$, valid in
the infinite population limit.

\subsection{A mean-field analysis gives a relation between the
  variance of the advantage distribution and the adaptation speed.}

\begin{figure}[htb]
  \includegraphics[width=0.46\textwidth]{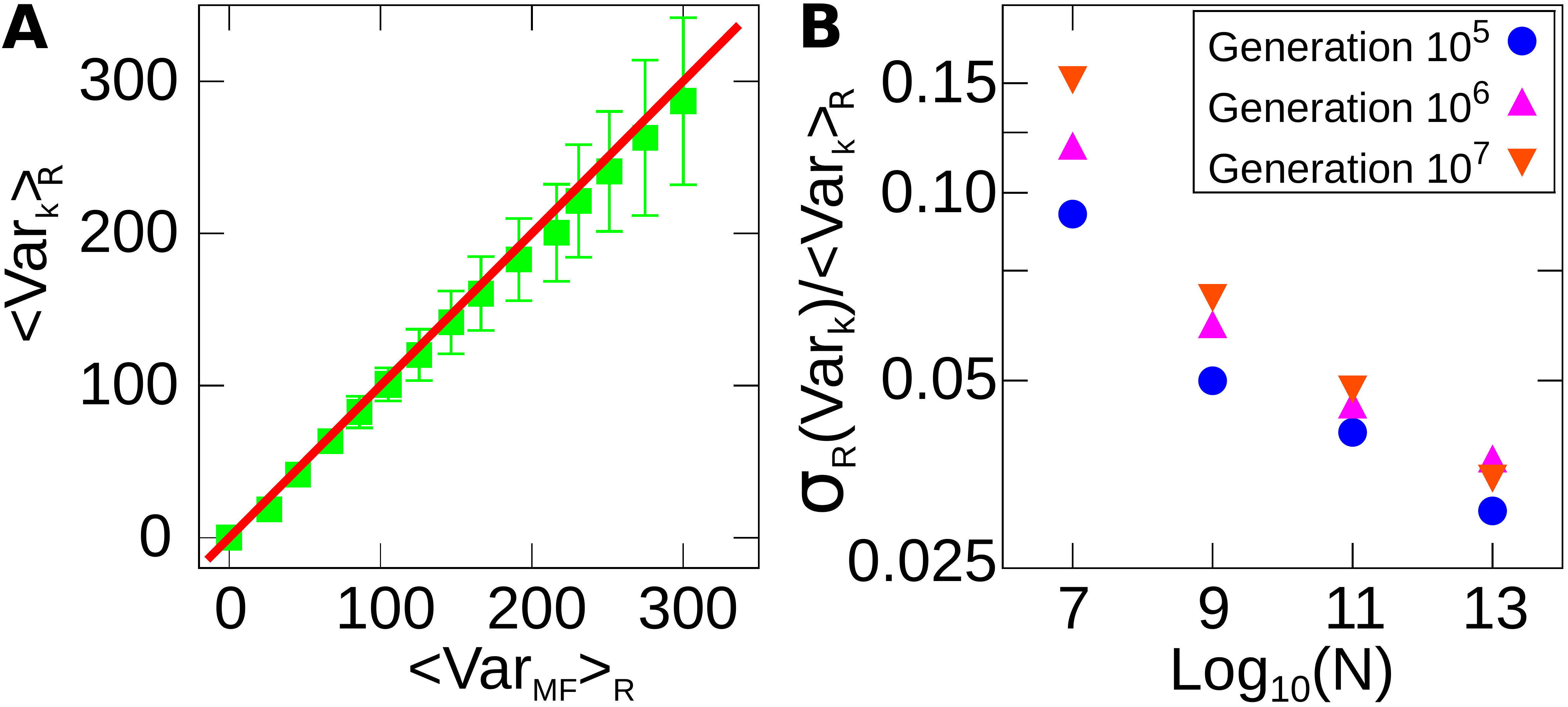}
    \caption{(Color online) \textbf{The mean field approximation
        captures a relation between $v_k$ and the width of the fitness
        class distribution, which is valid at intermediate times for
        moderate $N$ and until longer times for large population
        sizes.} (A) Simulated variance of the mutation classes
      distribution, shown as a function of the expected variance from
      the mean-field estimate ($\mathrm{Var}_{MF}= (v_k -U_b)\langle k
      \rangle^{1-\alpha}(s_0 \alpha)^{-1}$, see
      Eq~\eqref{eq:fisher_maria_theorem}). To avoid ambiguities,
      averages over realizations are indicated by a suffix $R$. The
      continuous red line represents the theoretical prediction
      $\mathrm{Var}_{MF}=\mathrm{Var}_k$.  The error bars (standard
      deviations over realizations, $\sigma_R(\mathrm{Var}_k)$) become
      larger with $\langle\mathrm{Var}_k\rangle_R$.  (B) While for
      increasing times $\sigma_R(\mathrm{Var}_k)$ diverges, the
      relative variability $\sigma_R(\mathrm{Var}_k)/\langle
      \mathrm{Var}_k\rangle_R$ over realizations decreases with
      increasing population size $N$, for any fixed time. This
      suggests that the infinite $N$ mean-field estimate is
      well-defined. Simulations are carried out using the parameters
      $s_0=0.5$, $\alpha=0.02$, $U_b=1 \cdot 10^{-3}$.  
      Population size in panel A is $N=10^7$.}\label{Fig_Meanfield}

\end{figure}

In the limit of infinite population, $N \to \infty$ the dynamics of
the model can be described using the following mean-field
equation~\cite{Park2010},
\begin{eqnarray}
  f(k,t)={}(1-U_b)f(k, t-1)\frac{w(k)}{\langle w \rangle_{(t-1)}}\:+ \nonumber\\
	   {}U_b\:
  f(k-1,t-1) \frac{w(k-1)}{\langle w \rangle_{(t-1)}} \ ,
\label{freq_meanfield}
\end{eqnarray}
where $f(k,t)$ is the frequency of individuals with $k$ beneficial
mutations at generation $t$, $w(k)=e^{s_0 g(k)}$, and $\langle w
\rangle_t=\sum_k w_k f(k,t)$  the mean fitness.

Multiplying Eq.~(\ref{freq_meanfield}) by $k$ and summing over $k$
gives the following expression for the dynamics of the mean number of
mutations $\langle k \rangle_t=\sum_k k f(k,t)$,
\begin{equation}
  \langle k \rangle(t+1)  = \frac{\langle k\:w\rangle(t)}{\langle
    w\rangle(t)} + U_b \ . 
  \label{k_meanfield}
\end{equation}
This expression can be further simplified assuming that the frequency
distribution is narrowly peaked around the mean (which travels in
time) and that it can be expressed as $f(k,t) \approx \delta(k;
\langle k \rangle(t))$.  In this case it can be easily verified that
$v_k = U_b$.

A different, more instructive, relation, which keeps into account the
width of $f(k,t)$ can be obtained starting from Eq.~\ref{k_meanfield},
and expanding $w(k)$ under the assumption that $D_k \equiv(k-\langle k
\rangle) \ll \langle k \rangle$, for every index $k$ of non-empty
classes. This assumption is verified by simulations (see
Fig.~\ref{fig_Var} in Appendix) and by further considerations on the
finite-$N$ width of the distribution given in the following sections.

Computing averages in Eq.~\eqref{k_meanfield} to first order in $D_k$,
and noticing that $\langle D_k \rangle = 0$ and $\langle D_k^2 \rangle
= \mathrm{Var}_k $, it is possible to obtain an expression for the
speed of accumulated mutations as a function of the variance of the
fitness class distribution.  Estimating $v_k$ as $ \frac{d\langle k
  \rangle}{dt} \approx \langle k \rangle(t+1)- \langle k \rangle(t) $
gives, for the case $g(k) = k^{\alpha}$
\begin{equation}
  \frac{d\langle k \rangle}{dt} \approx  
  s_0 \alpha \langle k\rangle^{\alpha-1}(t)
  \mathrm{Var}_k (t) + U_b  \ .
  \label{eq:fisher_maria_theorem}
\end{equation}
According to this equation, $v_k$ is driven by two terms, the increase
of mutations due to the beneficial mutation rate $U_b$ and the
selection of individuals with larger fitness.  This result is somewhat
reminiscent of Fisher's fundamental theorem and the Guess relation,
which relate the speed of adaptation to the variance of the fitness.
In this case, the speed of accumulation of successive mutations is
related to the width of the mutation class histogram, but rescaled by
the factor $\langle k^{\alpha-1} \rangle(t)$, which decreases with
time. For the non-epistatic case ($\alpha=1$) one recovers the more
usual linear proportionality, since the advantage is linear in the
mutation class index~\cite{Park2010}.

The mean-field limit of the model with $\alpha=1$ has been previously
addressed by Park et al~\cite{Park2010} with a moment generating
function approach.  In particular, they estimated the distribution
variance as $\mathrm{Var}_k\simeq \frac{1-U_b}{s_0}$, which
substituted in Eq.~\ref{eq:fisher_maria_theorem} gives precisely their
expression for the speed $v_k\simeq 1$ (in the limit of small $U_b$)
suggesting that our result is a consistent generalization to the
epistatic case.

Therefore, for $\alpha=1$ the speed of adaptation does not depend on
the mutation rate for infinite populations.  On the other hand, for
diminishing returns, the increase in the width of the distribution of
$k$, does not compensate for the term $\langle k^{\alpha-1}\rangle$
(which tends to 0), and, for long times,
Eq.~\eqref{eq:fisher_maria_theorem} predicts the limit velocity $v_k =
U_b$, as observed in simulations (Fig.~\ref{fig_KWV}C).
Since the increase in the width of the distribution depends on the
size of the population and on the advantage $s_0 g(k)$, the unphysical
limit velocity $v_k \approx U_b$ is approached with different laws
depending on the specification of $g(k)$.

It is necessary to discuss under which conditions this mean-field
limit can be considered a valid estimate of the typical behavior of a
realization.
At finite $N$, simulated data are in good accordance for intermediate
time with the mean field estimate for the variance $\mathrm{Var}_k$ of
the mutation class distribution
(Fig.~\ref{Fig_Meanfield}). Additionally, $\mathrm{Var}_k/\langle
k\rangle$ decreases quickly with time (see Fig.~\ref{fig_Var} in Appendix),
 justifying the assumption of small
$D_k/\langle k\rangle$ (since it is expected that $ |D_k| \lessapprox
(\mathrm{Var}_k)^{1/2} $ for every $k$).
However, the variability of $\mathrm{Var}_k$ over different
realizations increases quickly. 
In order to verify whether these fluctuations are well-behaved, we
have evaluated $\gamma=
\sigma_R\left(\mathrm{Var}_k\right)/\langle\mathrm{Var}_k\rangle_R$, where
$\mathrm{Var}_k$ indicates the variance of the mutation class
distribution in a single realization (roughly analogous to $L_k^2$), and the
suffix $R$ indicates averages over realizations. Specifically,
$\langle x\rangle_R$ indicates the average of the quantity $x$ over
different realizations, while $\sigma_R (x)$ is its standard
deviation. Therefore, $\gamma$, plotted in Fig.~\ref{Fig_Meanfield}B
as a function of $N$, represents the relative variability over the
realizations of the variance $\mathrm{Var}_k$ of the distribution.  
For any fixed time, this quantity decreases with $N$, suggesting that
the mean-field limit is well-defined for infinite populations.
Conversely, fixing $N$ and increasing $t$, $\gamma$ appears to reach
finite values, hence $\mathrm{Var}_k$ (and hence $v_k$) seems to be
non-self averaging in time.
This indicates that a mean-field description of the population
dynamics might be appropriate for longer time-scales at increasing
population size.

Note that these questions are superfluous for any empirical purpose of
the model, since the infinite $N$ mean-field limit does not describe
correctly the finite $N$ dynamics already for
$\alpha=1$~\cite{Park2010}, and the time scale of any experiment is
orders of magnitude smaller than those considered here.  ``Real''
evolutionary time scales are longer, but presumably subject to
changing conditions, and therefore beyond the scopes of this model.
However, the infinite $N$ limit is instructive on purely theoretical
grounds, for understanding the main mechanisms driving the model.

To sum up, the mean-field regime predicts that the
infinite-population/infinite-time speed for the mutation classes is
$U_b$.
At finite $N$, the long-time averages for the number of mutations $k$
and the associated speed $v_k$ agree with the predictions of the
mean-field regime, but are characterized by strong fluctuations.
Finally, an extended expression accounting for the finite width
$\mathrm{Var}_k$ of the class $f(k,t)$ relates $v_k$ to the width
itself, and is verified by simulations. In this long-time regime, a
number of assumptions of the model break down, and more complex models
need to be employed.
For practical purposes, neither the infinite population size nor the
infinite time limit are relevant, as even in long-term experiments the
number of generations of interest is typically sufficiently low.
This will be discussed in a later section, when the model parameters
are matched with empirical data from two laboratory evolution
experiments. The next sections discuss in greater detail the behavior
of the model at finite $N$.

\subsection{At finite population size, the advantage and
  fitness-class distributions remain Gaussian, but their width is
  unsteady, with opposed trends.}

The large-$N$ long-time phenomenology described so far and captured in
terms of mean-field estimates is not very relevant empirically, given
the experimental range of times and population sizes.  
In the following we offer a more detailed summary of the behavior of
finite-size populations at intermediate times (i.e. on the relevant
experimental time scale $\approx 10^2-10^4$ generations, with $\langle
k \rangle \approx 10-10^2$).

As already illustrated, the frequency distributions of the mutation and
advantage classes are approximatively Gaussian for a constant advantage
model.  Direct simulations of the diminishing return model show that
this still holds (Fig.~\ref{Fig_Gaussiane}A).
Given the discrete nature of these distributions, their widths are
well represented by the distances $L_k$ and $L_s$ of the foremost bin
from the average (shown in Fig.~\ref{fig1}A).
While in a fixed advantage model these distances, in terms of both
mutation and advantage classes, between the edge and the mean are
constant, for a diminishing return model $L_k$ is an increasing
function of the mean number of mutation classes $\langle k\rangle$,
while $L_s$ decreases with $\langle k\rangle$
(Fig.~\ref{Fig_Gaussiane}B).

The model without epistasis (and the mean-field estimates) make us
expect that the widths of these histograms are related to the speed of 
adaptation and consequently of mutation accumulation.
The instantaneous velocities of the two distributions $v_k$ and $v_s$
both decrease with $\langle k \rangle$. 

Qualitatively, one expects that, with changing mean number of
mutations, the two speeds are connected by the decay of the advantage
between $k$ and $k+1$. However, the relation between them is also
affected by the change in the width of the distributions with
increasing $\langle k \rangle$.
Indeed $v_s$ decreases more rapidly and, in the long-time limit, even
if $v_k > 0$, it vanishes.  This implies the existence of a large
number of sub-populations with different $k$ but essentially the same
fitness, which leads to the effectively neutral behavior discussed in
the previous section.
	Since the advantage decreases with time, and the distributions change
width, it is interesting to ask under which conditions the
biologically relevant regimes are preserved. Specifically, the
condition of multiple concurrent mutations could be lost over time.

As discussed in the Introduction, the clonal interference condition
can be roughly defined (neglecting irrelevant numerical factors) 
by the scaling relation $NU_b ln(N\Delta s_k)
\gg 1$, where $\Delta s_k = s_0\left(g(k_\mathrm{max}) -
  g(\langle k\rangle)\right)$ is the advantage of the fittest class. 
	This condition is satisfied when the population size and the edge
advantage $\Delta s_k$ are large enough, i.e. when $N \Delta s \gg 1$
and $N U_b \gg 1$.  Considering a fixed advantage model, for typical
values of experimental population size ($N\approx 10^6-10^{10}$) and
advantage ($s_0 \approx 10^{-2}$), these inequalities are always
satisfied, for $U_b$ estimated in the range $10^{-3}-10^{-10}$
mutations per genome per generation.
However, when the contribution of a newly arising mutation becomes too
small, and the advantage in fitness of the edge ($\Delta s_k$) is
smaller than $1/N$, the clonal interference hypothesis breaks
down. Note that this implies also an establishment size larger than
$N$.  These considerations confirm that, while the long-time limit of
this model is biologically implausible, the experimentally relevant
parameter regime should be captured correctly by our theoretical
framework.

We will now attempt to generalize the analytical estimates for the
finite-$N$ adaptation speed available for the multiple mutation model
to the case of diminishing returns.
These ``stochastic edge'' estimates are based on the hypothesis that
the only class subjected to substantial stochastic effects is the
fittest one, i.e. that $\Delta s \gtrsim U_b$~\cite{Trav_wave,
  Desai2007}.  When $\Delta s \approx U_b$ the single stochastic class
approximation fails, and the dynamics is more complex.
Supposing that $s_0 = 5\cdot 10^{-1}$, $\alpha = 0.02$ (this is a much
stronger epistatic effect than the one we estimate from experimental
data, see the following) and that $L_k \approx 50$ (of the order of
the values obtained from simulations with this parameter set and
population size $N=10^7-10^{13}$), the advantage becomes close to a
beneficial mutation rate of $U_b \approx 10^{-3}$ (which can be
considered very large~\cite{Hegreness2006,Perfeito2007}), for $\langle
k\rangle \approx 5\cdot10^2$.
This exceeds the interesting experimental range of $\langle k\rangle$
($10^1-10^2$), suggesting that for the relevant range we can always
suppose that the stochastic edge approximation is valid.

\subsection{Analytical estimate for the adaptation speed at finite $N$. }
\label{Analit_estimates}

\begin{figure*}[htbp]
  \includegraphics[width=0.85\textwidth]{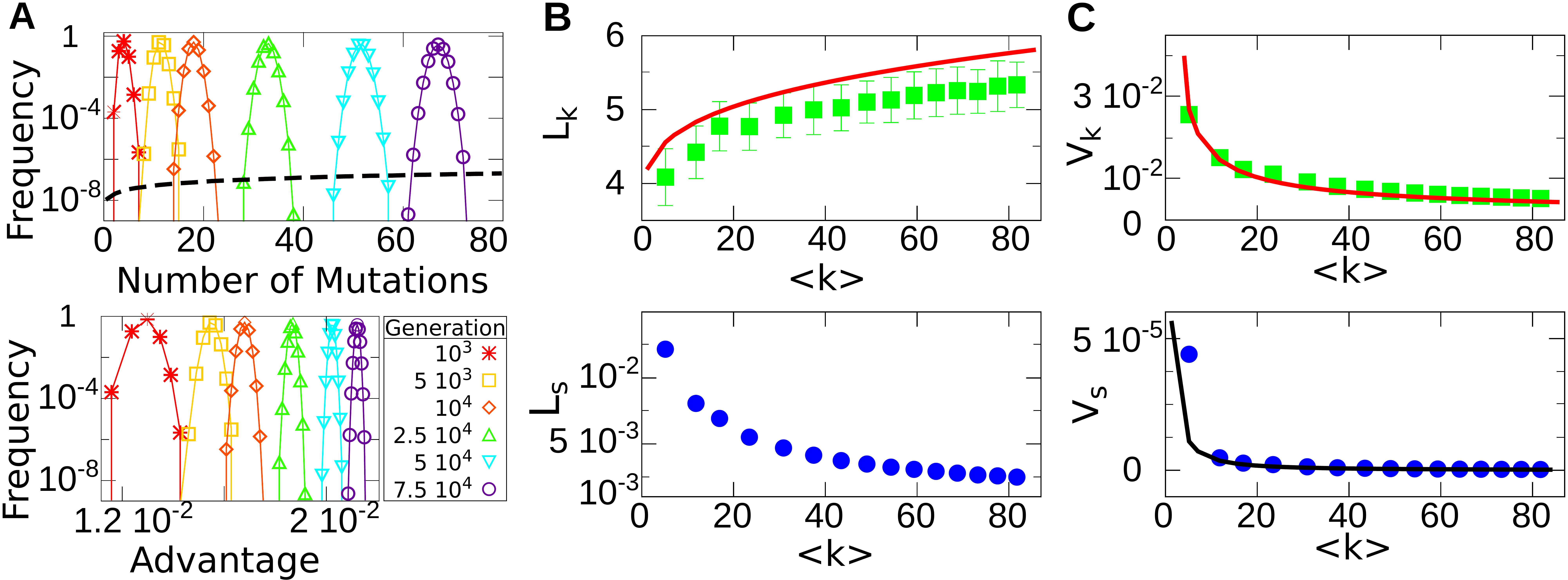} 
  \caption{ (Color online) \textbf{The histograms of fitness advantage
      and mutation classes have nearly Gaussian forms. While
      adaptation slows down, the latter histogram expands while the
      former becomes increasingly peaked.}  (A) Histograms of the
    mutation classes (top) and advantage classes (bottom) obtained
    from simulations averaging over 200 realizations at different
    generations (different symbols, see legend). The parabolic form in
    the semi-log plot indicates that they are approximately
    Gaussian. The establishment size is represented as a dashed line
    in the top panel. Solid lines connecting the symbols are guides
    to the eye.
    (B) Simulated data for the widths of the mutation class
    histogram $L_k$ (green squares, top) and of the fitness advantage
    histogram, $L_s$ (blue circles, bottom), plotted as a function of
    the mean number of mutations $\langle k\rangle$. 
    The continuous line represents 
    the theoretical estimates of the width
    (see Eq.~\eqref{eq:L_zero_ord}).
    (C) Plots of the speed of mutation accumulation $v_k$ (top, green squares)
    and of adaptation $v_s$ (bottom, blue circles), as a function of
    the mean number 
    of mutations $\langle k\rangle$.  Continuous lines are the corresponding
    theoretical estimates (see Eq.\eqref{eq:v_k_stima1} and
    \eqref{eq:v_s_stima1}). Note 
    that since $\tau_k$ depends logarithmically on $L_k$, the
    estimates for both speeds are in satisfactory agreement
    with the simulated data even if $L_k$ is approximated more roughly.
	The parameters used in the simulations are $N=10^9, \
    U_b=6\cdot10^{-6}, \ s_0=0.1,\ \alpha=0.2 $, compatible with
    those estimated (see Sec.\ref{Param_estim}). Averages are
    performed over 200 realizations.
  }
  \label{Fig_Gaussiane}
\end{figure*}

Since we have verified that the advantage and mutation class
histograms are both nearly Gaussian (but not stationary in width), it
is possible to attempt a generalization of the estimates applied for
the standard multiple mutation model, Eqs.~\eqref{eq:DF1} and
\eqref{eq:DF2}.  Supposing a slow increase of the width of the
distribution with $k$ and assuming that the width of the histogram
is stable while the new class is being established, the mean of the
distribution moves from $\langle k \rangle_t$ to $\langle k \rangle_t
+1$ during the establishment time $\tau_k$ and $L_{k}\approx L_{k+1}$.
For $\langle k \rangle \gg L_k > 1$ (and $\langle k \rangle$ not too
high due to the condition of sufficiently large $\Delta s$) the
advantage of the edge respect to the mean class is $\Delta s_k= s_0(k^{\alpha}-
\langle k \rangle^\alpha) \approx s_0 \alpha L_k k^{\alpha -1}$.

Thus, imposing the condition that one new mutation class is
established at the edge of the histogram gives 
 
\begin{eqnarray}
  1=  \int_{0}^{\tau_k}    {}\mathrm{d}t  \left(
    \frac{U_b}{2 s_0 \alpha L_k k^{\alpha -1}} e^{s_0 \alpha(L_k -1) k^{\alpha -1}t}
  \right)\nonumber\\
\left(2 s_0 \alpha L_{k+1} (k+1)^{\alpha -1}  \right)
 \ .
\label{eq:STIMA1}
\end{eqnarray}

As in Eq.~\eqref{eq:DF1}, the last term in this integral is the
establishment probability of a new fittest class, while the first is
the rate of beneficial mutations from the previous fitness class,
which is born with size $\frac{1}{2s_0 \alpha L_k k^{\alpha-1}}$ and
grows exponentially. Note that imposing $\alpha =1$,
Eq.~\eqref{eq:STIMA1} becomes Eq.~\eqref{eq:DF1}, recovering the
standard multiple-mutation model.

An estimate of $\tau_k$ can be obtained integrating the above
expression
\begin{equation}
  1= \frac{U_b}{s_0 \alpha(L_{k} -1)} \frac{(k+1)^{\alpha -1}}{k^{2(\alpha -1)}}
  [e^{s_0 (L_k -1) k^{\alpha -1}\tau_k}  -1] , 
\label{eq:DF1_tau_boundary}
\end{equation}
where we used the approximation $L_{k}\approx L_{k+1}$.  Under the
assumption that $k$ is not too large (i.e. the advantage of the
fittest class is sufficiently high) the contribution of the
integration boundary $t=0$ can be neglected and the expression of
$\tau_k$ is the equivalent of Eq.~\eqref{eq:DF1_tau} for the
non-epistatic case,
\begin{equation}
  \tau_k = \frac{1}{s_0 \alpha (L_{k} -1) k^{\alpha -1}} \log\left(
   \frac{s_0 \alpha(L_k -1) k^{2(\alpha-1)}}{U_b (k+1)^{\alpha -1}}\right) \ .
  \label{eq:STIMA1_tau}
\end{equation}

The above expression can be further simplified assuming that both $k$
and $L_k$ are large enough so that $k+1 \simeq k $ and $L_k -1\simeq
L_k $ and neglecting the logarithmic term in $L_k$ leading to the
following expression
\begin{equation}
  \tau_k = \frac{1}{s_0 \alpha L_{k} k^{\alpha -1}} \log \left( 
  \frac{s_0 \alpha  k^{(\alpha-1)}}{U_b}\right) \ .
  \label{eq:STIMA1_tau_DF}
\end{equation}
This indicates that for intermediate $k$, the estimate of the standard
multiple mutation model is valid provided the diminishing return
advantage function $s_0 k^{\alpha}$ is substituted to the constant
advantage. For sufficiently large $k$, expansion of the exponential in
Eq~\eqref{eq:DF1_tau_boundary} gives $\tau_k \sim 1/U_b$,
compatible with the mean-field result.

The time $\tau_k$ is related to the instantaneous speed of the
mutation class histogram $v_k$.  The speed $v_s$, can be obtained
knowing that during the time the mutation class histogram travels by
one class, the advantage histogram has to move by the relative fitness
between the newly added fittest class and the previous one, $s_0
\alpha k^{\alpha -1}$. Using the simplified
Eq.~\eqref{eq:STIMA1_tau_DF} (which neglects logarithmic terms in
$L_k$), one obtains the speed
\begin{equation}
  v_k = \frac{s_0 \alpha L_k  k^{(\alpha -1)}}{\log(\frac{s_0 \alpha
      k^{\alpha -1}}{U_b})}  
\label{eq:v_k_stima1}
\end{equation}
and
\begin{equation}
  v_s = \frac{s_0 \alpha k^{\alpha -1}}{\tau_k} =  \frac{s_0^2 \alpha^2 L_k
  k^{2(\alpha -1)}}{\log(\frac{s_0 \alpha k^{\alpha -1}}{U_b})} \ .
\label{eq:v_s_stima1}
\end{equation}

The second part of the estimate involves the normalization
condition~\eqref{eq:DF2}. Assuming (as in the standard estimate) that
the largest term of the histogram dominates, we need to evaluate the
time $\tau_k'$ necessary for the fittest class to become the class
with mean advantage, whose size is order $N/2$.
If the fittest class has $k$ mutations, its establishment size is
$\frac{1}{2 s_0 \alpha L_k k^{\alpha -1}}$.  Its growth will be
roughly exponential, with a rate that decreases while it gets closer
to the mean. We estimate its growth by its mean growth rate during the
time $\tau_k'$. Immediately after establishment, its relative growth
rate will be $s_0 \alpha L_k k^{\alpha -1} $, while its rate will tend
to zero when it gets close to the mean.  Thus, on average, we can
assume that it grows exponentially with rate $ \frac{s_0 \alpha L_k
  k^{\alpha -1}}{2}$.

This argument leads to the equation 
\begin{equation}
  N/2 \approx \frac{1}{2 s_0 \alpha L_k k^{\alpha -1}} e^{\frac{
      s_0\alpha L_k k^{\alpha -1}}{2} \tau'_k} 
\label{eq:STIMA2}
\end{equation}
which corresponds to Eq.~\eqref{eq:DF2}, and implies the equivalent of
Eq.~(\ref{eq:DF2_tau}),
\begin{equation}
  \tau_k'= \frac{2}{s_0 \alpha L_k k^{\alpha -1}} 
  \log\left( N s_0 \alpha L_k k^{\alpha -1}\right) \ .
\end{equation}

In order to estimate $v_s$, we need to determine how much the
histogram of fitness advantage has progressed during the time
$\tau'_k$ from the establishment of the $k$-th mutation class. We
assume that during this time $L_k$ is roughly constant, so that after
time $\tau'_k$, $k+L_k$ mutations are established, and the advantage
of the edge has reached $ s_0 \alpha L_k (k +L_k)^{\alpha -1} $.

This allows to estimate $v_s$ as the advantage gained divided by the
time $\tau'_k$, i.e.
\begin{equation}
  v_s = \frac{(s_0 \alpha L_k)^2}{2}  
   \frac{k^{\alpha -1} (k+L_k)^{\alpha -1}}{ \log(N s_0 \alpha L_k k^{\alpha -1})} 
        \label{eq:v_s_stima2}
\end{equation}

Eq.~\eqref{eq:v_s_stima1} and~\eqref{eq:v_s_stima2} together allow to
determine $L_k$, which can subsequently be used to obtain the speed of
adaptation $v_s$, or of fixed mutations $v_k$, using e.g.
Eq.~\eqref{eq:v_s_stima1} and~\eqref{eq:v_k_stima1}.
Assuming that $k + L_k \simeq k$ and
neglecting the logarithmic corrections in $L_k$, as made to obtain
\eqref{eq:v_s_stima2}, we can  obtain the
following closed expression for $L_k$,
\begin{equation}
L_k = \frac{2 \log( N s_0 \alpha k^{\alpha -1})} 
	{\log \left( \frac{s_0\alpha k^{\alpha -1}}{U_b}\right)}\ .
       \label{eq:L_zero_ord}
\end{equation}

Comparison with simulated data shows that the above expression for
$L_k$, despite the rough approximations made, is a reasonably good
estimate of the width of the distribution
(Fig.~\ref{Fig_Gaussiane}B). In particular, the speed of adaptation
and mutation accumulation (Fig.~\ref{Fig_Gaussiane}C) are well
captured by our analytical description.
We should stress again that these expressions can be considered valid 
for intermediate values of $k$, as explained above.
For large $k$, the integration boundary in
Eq.~(\ref{eq:DF1_tau_boundary}) cannot be neglected, and it can be
verified that $v_k=1/\tau_k$ tends to a constant, $U_b$ with the approximations
taken, restoring the correct mean-field limit.


\subsection{Parameter matching of the model with data from laboratory
evolution experiments.}\label{Param_estim}

Having explored some of the main features of the diminishing return
model using theoretical arguments and simulations, we now proceed to
compare it to experimental data. We have made very clear that this
model is a very crude description of any realistic situation. However,
its advantage is that it is simple and based on few parameters, so
that estimating some of the model's fundamental parameters from data
is a relatively simple task, which could be
instructive~\cite{Desai2007a}.

We developed a parameter matching procedure that produces an estimate
of the beneficial mutation rate. This estimate depends on the specific
advantage model chosen, but also allows to select models according to
how well they resemble the data. Comparing the simulated dynamics with
the experimental one, it is possible, for each specific data set, to
determine an optimal functional form of the advantage $g(k)$, thus
narrowing down the estimate of $U_b$.

We analyzed fitness/mutations data from two laboratory evolution
experiments using the three different variants of the diminishing
return model described in Appendix~\ref{appendix:modelli}.  The power
law model, where the advantage is described by $s_0g(k)=s_0 k^\alpha$,
is the main case presented in the former sections. The advantage
functions of other two models considered have a logarithmic ($g(k)=
\ln(k+1)$) and an exponential ($g(k)= \frac{1-q^k}{1-q}$) dependence
on $k$.  These functions can be derived, with some approximations, as
partial sum of the harmonic and geometric series, which allows to
produce simple expressions of the 
antagonistic effect of each added mutation. 
These two model variants share most of the
qualitative features of the main formulation.

\subsubsection{Experiments analyzed}

We applied our model to two experiments of controlled evolution. The
first are the first 20000 generations of the well-known
``\emph{E. coli} long-term evolution''
experiment~\cite{Barrick2009,LTEE_Ecoli, Lenski1991}, while the second
is a chemostat experiment performed by some of the
authors~\cite{Thomen2012}.

The two experiments concern two bacteria (respectively
\emph{Escherichia coli} and \emph{Acinetobacter baylyi}), were
performed using distinct propagation techniques (serial dilution in
batch and chemostat respectively) and also their duration is quite
different both in terms of generations ($2\cdot10^4$ and $3\cdot10^3$)
and experimental time ($\approx 10$ years and $\approx 4$ months).
Another remarkable difference between the two experiments is the
population size, which varies every day by two orders of magnitude
(between $\approx 5\cdot 10^6-10^8$) for the serial dilution
experiment while it is very large and approximately fixed in the
\emph{A. baylyi} experiment ($\approx 3\cdot10^{10}$).
The two experiments share interesting features suggesting that a
diminishing return model might be applicable to describe the
evolutionary dynamics of the populations.  Firstly, their duration in
terms of generations is long enough to observe a deceleration
of fitness increase~\cite{Kryazhimskiy2009}.  Secondly, the large
(effective) population size suggests that the clonal interference
regime might be relevant in these experiments. The simultaneous
presence of different genotypes within the population has been
verified in both
experiments~\cite{Thomen2012,Elena1997}.
Moreover, the decrease of the beneficial effect of the first five
fixed mutations in the \emph{E. coli} serial dilution experiment has
been recently demonstrated~\cite{Khan2011}. The effect is more complex
than the description of diminishing returns given here. However, as
suggested by the authors, one can surmise that a simplified model
including epistatic interactions might be useful to roughly describe
this phenomenon.

A detailed description of the data and the choices made for the
analysis procedure is given in
Appendix~\ref{appendix:experiment}. Here we give a brief description
of the main features of the two experiments.

The \emph{A. baylyi} experiment studied the population dynamics in a
chemostat using a minimal medium supply for about four months, at a
dilution rate $D\approx 0.7 h^{-1}$.  The use of chemostat
allowed to grow a large population ($N \approx 3\cdot10^{10}$)
under controlled conditions for a fairly long time.  Since
the number of individuals is large, it is expected that different
sub-populations will grow in parallel in clonal interference
regime. This has been confirmed by population sequencing
data~\cite{Thomen2012}.
Additionally, several single clones where isolated from samples
collected and frozen at different time. 

 The maximum growth rate
($\mu_{max}$) of 21 isolated clones and of the the original strain
introduced in the chemostat (wild type) has been measured in batch,
fitting the growth curve during the exponential phase.
  We considered the maximum growth rate measured in batch as indicative of
the fitness of the population into the chemostat, defining the
normalized fitness of the clone $i-th$ as
$w_{exp}(i)=e^{\mu_{max}(i)-\mu_{WT}}$.  A more detailed description
of this procedure is given in App.~\ref{appendix:experiment}.
Note that, since the reference fitness value is given by the ancestral
growth rate, $w_{exp}(\mathrm{WT})=1$.
A whole-genome sequencing was performed on two single clones isolated
at the end of the experiment (AB2800b and AB2800a) and on the
ancestral strain that served as a reference for identifying mutations
in evolved clones.  A total of 11 mutations were
detected into the evolved clones, eight of them in common between the
two.  Additional sequencing has been performed in the remaining clones
selected at different generations on PCR fragments encompassing the
mutated loci identified in the two end-point clones in order to
reconstruct a sketch of the history (or the genealogy) of the mutation
appearances.

The \emph{E.~coli} long-term evolution experiment concerns twelve
\emph{E.~coli} populations evolved in parallel for about $5\cdot10^4$
generations in batch. Serial dilution was performed daily, allowing
$\approx 6.64$ generations each day and an effective population size of
$\approx 2\cdot10^7$ individuals.  We used the mutations and fitness
data of the population designated Ara-1 referred to the first $20000$
generations, as given in ref.~\cite{Barrick2009}. 
The relative log-fitness of the ancestor and evolved populations(that
is essentially the ratio between the Malthusian parameters,
$\varphi_{i}=\mu_i/\mu_0$) was measured through competition
experiments (see ref.~\cite{deVisser2002}) every $1000$ generations.
We define in this case the mean fitness of the $i-th$ sub-population
as $w_{exp}(i)=e^{(\varphi_{i}-1)\bar{\mu}}$, where
$\bar{\mu}=\ln(2)$ is the approximate mean growth rate (see
ref.~\cite{Lenski1991} and Appendix~\ref{appendix:experiment}). The
fitness of the reference strain is again $w_{exp}=1$.
Genome sequencing was performed on samples from generation $2000$,
$5000$, $10000$ $15000$ and $20000$ as well as on the ancestor. A
total of $45$ mutations were found in the most evolved strain, most of
which were stable in later clones~\cite{Barrick2009}.

For all the analyzed clones in both experiments, it is possible to
associate fitness values with numbers of mutations, and thus bridge
with the parameter $k$ in our model.  For the \emph{A.~baylyi}
experiment the estimate of the number of accumulated mutations are
inferred from the genealogy, and thus have to be considered as a lower
bound for all the clones, except for the ancestor and the two fully
sequenced clones where they are measured directly. However, the
indications of $k$ from the population sequencing data are compatible
with the inferred values of $k$~\cite{Thomen2012}.

\subsubsection{Matching procedure and estimate of $U_b$}

\begin{figure}[htbp]
  \includegraphics[width=0.45\textwidth]{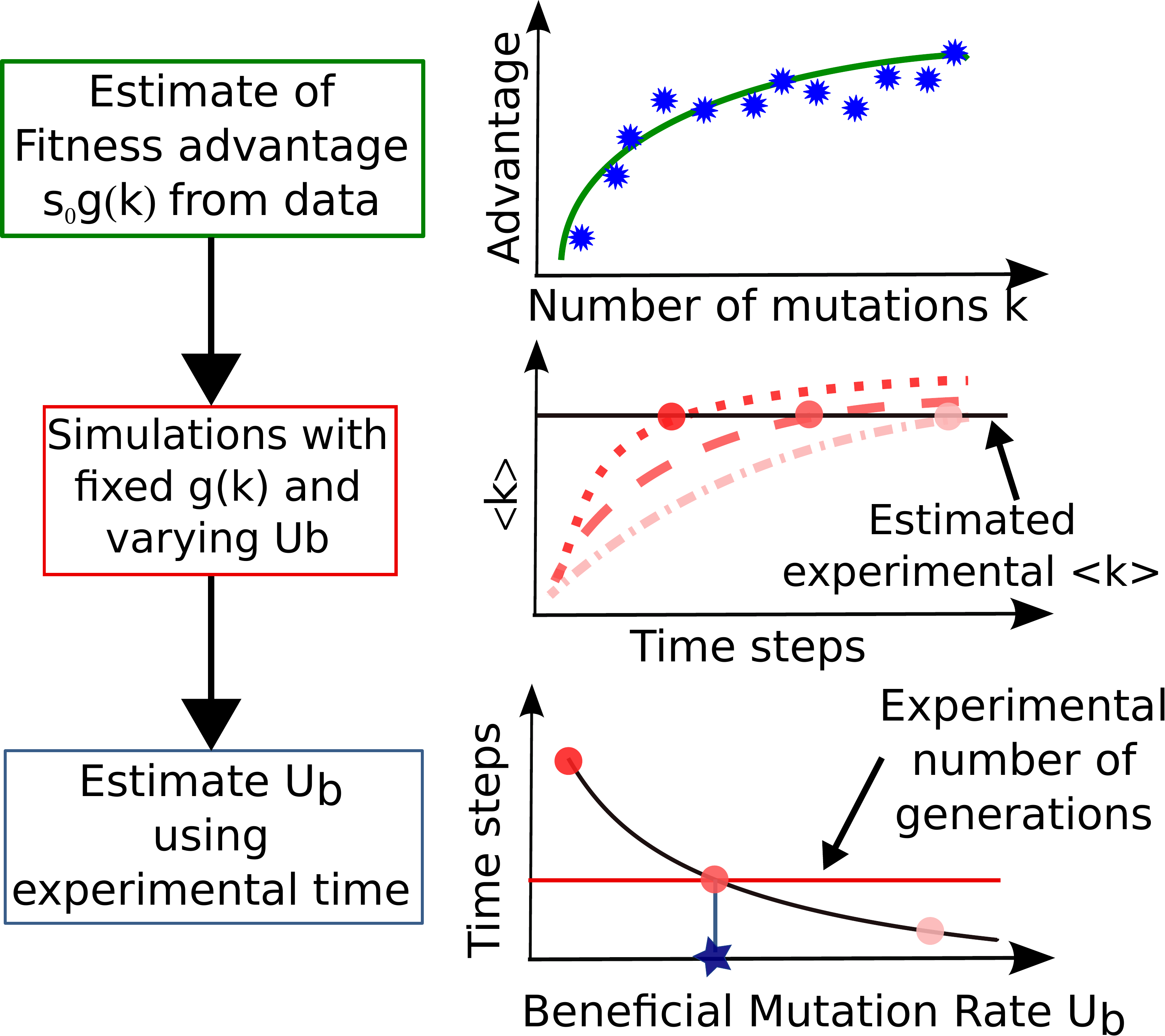} 
  \caption{(Color online) \textbf{Sketch of the parameter matching procedure
    leading to an estimate of the beneficial mutation rate from
    experimental fitness data}. Top: the experimental data for fitness
    as a function of the number of mutation (blue symbols) allow to
    estimate the advantage fitness function $s_0 g(k)$ from a fit
    (continuous green line).
    Middle: simulations are run using the estimate advantage function
    for different values of $U_b$, which remains undetermined.  This
    leads to different predicted dynamics for the number of acquired
    mutations (dotted,dashed, and dashed-dotted lines) and the fitness
    increase in time.  These predictions can be matched with the
    experiment to estimate $U_b$. In particular, the predicted number
     of time steps necessary to reach  
    the final experimental number of mutations  varies with the beneficial
    mutation rate (filled circles).   
    Bottom: The estimate of $U_b$ is obtained matching the predicted
     final time with the experimental one.  The best value of
    $U_b$ (blue star) is obtained when the predicted number of time
    steps necessary to reach the final number of mutations (determined
    in the experiment) corresponds to the number of experimental
    generations (horizontal continuous red line).
    We applied this procedure to the three model variants for the
    diminishing return described in the main text.}
  \label{fig:match}
\end{figure}

\begin{figure}[htbp]
  \includegraphics[width=0.43\textwidth]{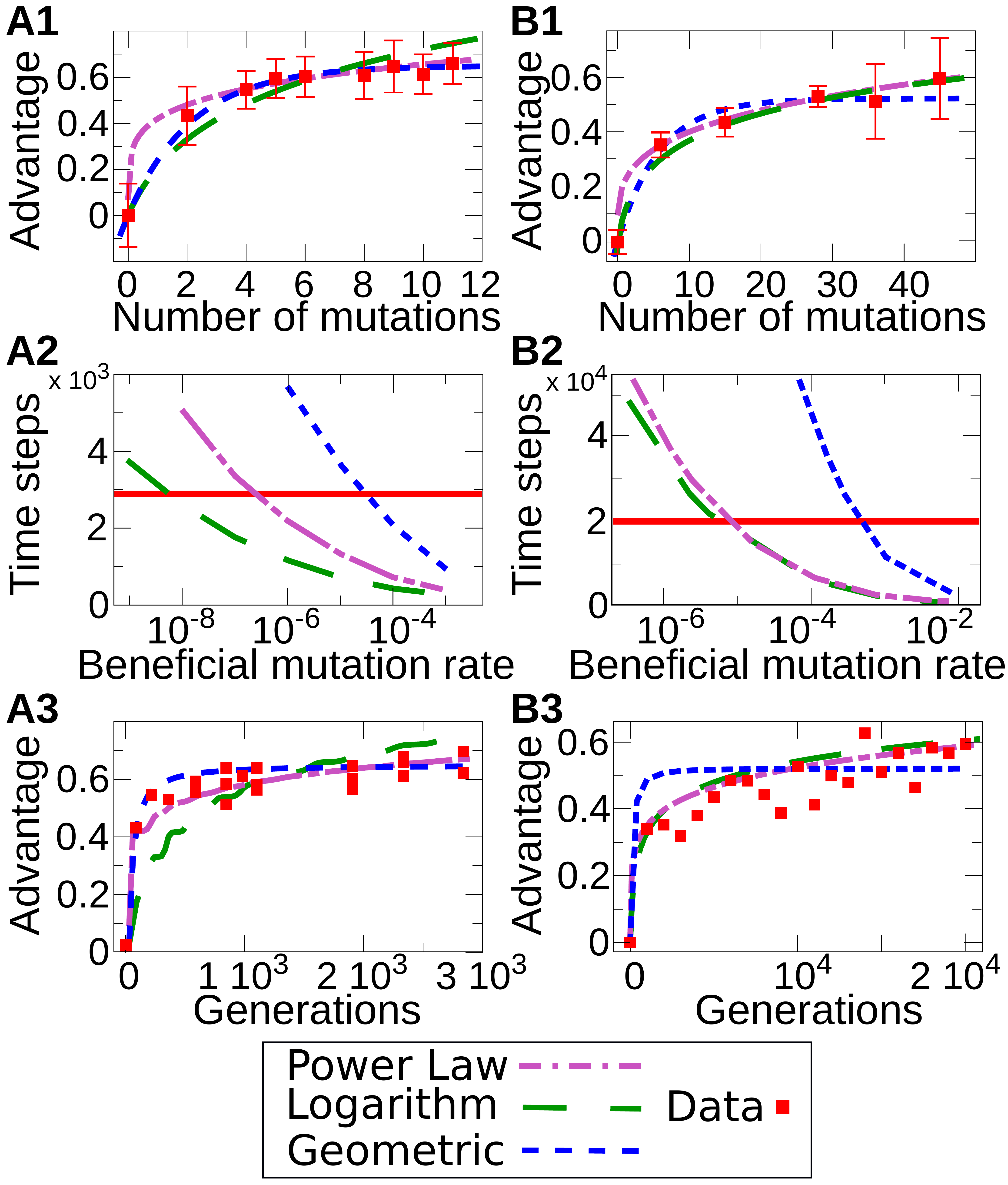} 
  \caption{(Color online) \textbf{Estimate of the beneficial mutation
      rate from different experiments.} 
    Estimate of the beneficial mutation rate 
    using data from Jezequel \emph{et al.}~\cite{Thomen2012} (left panels) and
    Barrick \emph{et al.}~\cite{Barrick2009} (right panels).  
    Top: estimate of the fitness advantage  function $s_0 g(k)$ from
    data of advantage as a function of mutation number (red symbols),
    as described 
    in the top panel of Fig.~\ref{fig:match}.  The different lines
    correspond to the three model variants for the diminishing-return
    advantage (power-law dot-dashed purple lines, logarithmic
    long-dashed green lines, 
    geometric blue dashed lines, as in legend). For sake of
    simplicity for experiment \emph{A.~baylyi} (panel A1) for each
    $k$ only the mean value of the advantage is shown.
    Complete data are reported in ref.~\cite{Thomen2012}.
    Middle: estimates of $U_b$, from different model variants,
    obtained matching the time for which
     $\langle k \rangle(t_{exp})=\langle k \rangle_{exp}$
    predicted by simulations with a given 
     $s_0 g(k)$ to the experimental number of generations, as described 
     in the bottom panel of Fig.~\ref{fig:match} (the different line
     styles refer to the model variants as above). For  \emph{A.~baylyi}
     experiment we considered a total of $2850$ generations.
     Bottom: comparison of the performance of the three model
       variants with the estimated values of $U_b$. For the Jezequel
       \emph{et al.}  data (left panel), the power law and geometric
       variants are quite close, but power law model for the
       diminishing return gives the best agreement, especially
       considering the data relative to the number of acquired
       mutations as a function of time (not shown). This choice leads
       to the estimate $U_b\approx 10^{-6}-10^{-7}$. For the Barrick
       \emph{et al.} data, the logarithm and power-law models perform
       better, and give equivalent estimates of $U_b\approx
       10^{-5}$.}
  \label{fig:exp_match}
\end{figure}

Figure~\ref{fig:match} summarizes the parameter matching procedure.  The
first step finds the best-fitting parameters for each functional form
of the fitness advantage function $g(k)$.  This uses data on fitness
values and number of acquired mutations, using the definitions given
in the previous section.
Since the number of experimental points is low ($5$ for the
\emph{E.~coli} and $21$ for the \emph{A.~baylyi} experiment-
corresponding to $9$ different values of $k$),
it is possible to obtain good fits with different functional forms of
the advantage $s_0 g(k)$. The results of the fit using the three
different functional forms considered here (power law, logarithm and
geometric) is shown in the top panel of Fig.~\ref{fig:exp_match}.
In a second step of the procedure, simulations are repeated for a wide
range of values of $U_b$.  An estimate of this parameter can be
obtained using the mean number of mutations at the end of the two
experiments (Fig.~\ref{fig:match}). In our case, $\langle k \rangle =
11$ and $\langle k \rangle = 45$ for the \emph{A.~baylyi} experiment
and \emph{E.~coli} experiment respectively.  
In the model, for a fixed interval of time steps, the number of
accumulated mutations decreases significantly and monotonically with
$U_b$, and the number of time steps necessary to reach $\langle k \rangle
= 11$ and $\langle k \rangle = 45$ depends on $U_b$. Thus, for each
model there is a single value of the beneficial mutation rate that
verifies $\langle k \rangle(t_{exp})=\langle k \rangle_{exp}$ 
(Fig.~\ref{fig:match} and middle panel of Fig.~\ref{fig:exp_match}).
Note that we are referring to the estimated values of $k$ as mean
values assuming that the sequenced clones are
representative of the populations. For the \emph{A.~baylyi}
experiment, the uncertainty on the number of mutations present in the
clones is a relevant source of error. On the other side, even
if for the \emph{E.~coli} experiment the number of mutations
is exact, and referred to single clones, the fitness data we
 associated are population mean fitness
data, that can introduce an error in the estimate.

The procedure applied so far gives some estimated values of the
beneficial mutation rate, which however,depend
 on the advantage model, varying at this stage up to three
orders of magnitude. 
 Nevertheless, all the estimates are roughly in
the expected biological range ($10^{−8}-10^{−5}$, see
Ref.~\cite{Perfeito2007,Hegreness2006}).
The third step of the estimate procedure allows to select between
advantage model, and is based on the comparison of simulated and
experimentally measured dynamics for the fitness and the number of
mutations as a function of time.
Note that this is not trivially equivalent to the fitness as a
function of $k$, but contains the effects of the population dynamics
in presence of clonal interference, for which the model provides a
description.
Since different functional forms of the advantage imply a different
behavior of the population, one can check which of the advantage
functions gives the closest description to the experiments.
Thus, the experimental data of the advantage as a function of the time
can be used to discern between power law, logarithmic and geometric
model, ultimately selecting an estimate of $U_b$.
 
A qualitative comparison between data and simulations shows that for
the \emph{A.~baylyi} experiment the power law and geometric advantage
model (with parameters $s_0=0.420$, $\alpha=0.194$ and $s_0=0239$,
$q=0.631$ respectively) best describe the increase of the fitness with
time (bottom panel of Fig.~\ref{fig:exp_match}). Comparing the
increase of the number of mutations as a function of time of the two
models suggests that the power law model better resembles the data.
For this reason, we speculate that, despite of the large errors in the
fits, the power law model could perform better in this
parameter-estimation procedure and the preferred estimated value of
beneficial mutation rate is around the value $U_b\approx
3\cdot10^{-7}$.
In the \emph{E.~coli} experiment, the power-law and logarithm model
result describe the data best, and are roughly equivalent.  
The estimate parameters are $s_0=0.220$, $\alpha=0.258$ for the 
power law and $s_0=0.158$ for the logarithm model.
The estimates for the beneficial mutation rate are also essentially
equivalent (considering the errors connected to the procedure) for the
two preferred models $U_b \approx 1\cdot10^{-5}$ for the power law and
$U_b \approx 6\cdot10^{-6}$ for the logarithmic advantage model. Note
that for this step of the procedure the data used are much more
abundant, since we used all the fitness data measured every $1000$
generations.
A logarithmic increase of the fitness advantage for \emph{E.~coli}
long term evolution experiment is also suggested by a parallel
work~\cite{Wielgoss_2012}.

\section{Discussion and conclusions}

Different laboratory evolution experiments show a decrease of the
fitness advantage due to newly acquired mutations and a decrease of
the speed of evolution~\cite{Elena2003,Kryazhimskiy2009,Hindre2012}.
This effect can be explained in different ways, and accordingly
different models have been formulated in this context. For example,
the speed of evolution could decrease because beneficial mutations
with larger advantage fix sooner in the
population~\cite{Schiffels2011} and because the mutation rate or the
number of possible beneficial mutations decreases with
time~\cite{Trav_wave,Park2008}.
Another explanation, suggested by recent experimental
observations~\cite{Khan2011,Chou2011}, could be that increasing the
number of accumulated mutations, epistatic interactions cause a
decrease of their effect on the fitness. These different explanations
are not necessary mutually exclusive, and could be stratified in
actual laboratory evolution experiments~\cite{Tenaillon2012a}.

We considered a simplified model, using a minimal number of
parameters, which is a direct generalization of the multiple mutations
model at constant advantage, but describes diminishing
returns. Specifically, it is assumed that the selective advantage of
all individuals having $k$ beneficial mutations is identical, but
decreases with $k$.
While more complex and realistic descriptions applicable to laboratory
evolution exist an advantage of the approach taken here is that, as
the multiple mutation model, it depends on few parameters, and as we
have shown, in line of principle allows matching of these parameters
with data. 
On theoretical grounds, this class of models is interesting because
its dynamics is driven by rare events~\cite{Desai2007,Brunet2008}, and
because its behavior at finite $N$ is qualitatively distinct from its
mean-field behavior~\cite{Park2010}.

We have shown that basic phenomenology of the model entails a
sublinear decrease of the mean number of fixed mutations and a steeper
sublinear decrease of the mean advantage.
This is in qualitative agreement with previous results using a similar
model applicable in a regime where concurrent mutations do not
occur~\cite{Kryazhimskiy2009}.

The two evolutionary speeds are related to the width of the
distribution of coexisting advantage classes, and thus of coexisting
mutation classes. We showed how a theoretical mean-field argument
produces a relation between the speed of fixed mutations $v_k$ and the
second moment of the histogram of mutation classes, confirmed by
simulations.
In the limit case of constant advantage, this relation reduces to the
one previously derived in ref.~\cite{Park2010}, related to Fisher's
fundamental theorem.
Interestingly, simulations indicate that for any finite $N$, different
model realizations behave increasingly differently with time in terms
of both $v_k$ and width of the mutation class histogram.
This non-self-averaging property implies that even at intermediate
times, the behavior of a realization can be quite different from the
average.

Within a mean-field framework, we also derived arguments showing that
the mutation speed $d\langle k\rangle/dt$ approaches the beneficial
mutation rate $U_b$ for very large times, which corresponds to the
mean behavior observed in simulations at large times.
In fact, as the original multiple mutations model, the present model
does not include deleterious mutations. However, in the case of
diminishing returns the assumption is more delicate, because the
advantage of fixed mutations will decrease indefinitely, until no
beneficial mutation is able to fix. Empirically, before this happens,
beneficial mutations will be balanced by deleterious
ones~\cite{Rouzine2003}.
This makes the (effectively neutral) long-time limit of the model
unphysical.
We have discussed how this limitation should not affect time-scales
and parameter values that are relevant for current laboratory
evolution experiments focused on adaptation~\cite{Hindre2012}. A
generalization of the model that includes deleterious mutations is
possible and could be the subject of future investigations, since it
allows to formulate evolutionary questions related to the balance of
mutations of different kinds~\cite{Rouzine2003,Trav_wave}.

Finally, for finite population size $N$, we were able to define
through analytical arguments the regime where the stochastic edge
estimate of the adaptation and mutation speeds can be extended to the
case of diminishing returns.
In the simplest case it is possible to obtain closed expressions for
the mutation class and adaptation speed and for the supports $L_k$ and
$L_s$ of the mutation class and advantage histogram respectively.
These expressions are direct generalizations of those obtained in the
case of constant advantage~\cite{Desai2007,Brunet2008}, and compare
well with numerical simulations.
Once again, in presence of diminishing returns these expressions,
besides the usual limitations in terms of parameter values, have a
limited applicability in time, but are typically valid in the
experimental range.

The constant-advantage multiple mutation model has previously been
applied to short-term laboratory evolution
experiments~\cite{Desai2007a}. In those early stages, adaptation does
not slow down, and the assumption of constant advantage is justified,
allowing to use the model to estimate empirical parameters.
However, on longer but experimentally observable time scales, the
assumption breaks down, and arguably the closest extension of that
model that can be used to estimate parameters is the one described
here.

In order to show a proof-of-principle parameter-matching procedure for
the diminishing return model, we considered two different experimental
data sets from long-term evolution experiments, and defined a
procedure that can be used to infer the order of magnitude of the
beneficial mutation rate from genomic and fitness data, assuming the
model.
This procedure allows to estimate the beneficial mutation rate $U_b$
and also compare different functional forms of the decreasing
advantage functions against data, through simulations of the model.
The values obtained for the beneficial mutation rate fall within the
range of the available measurements~\cite{Perfeito2007,Hegreness2006},
on the order of $10^{-6}/10^{-5}$ mutations per genome per generation for the
\emph{E.~coli} long-term evolution experiment, and between $10^{-7}$
and $10^{-6}$ in the case of \emph{A.~baylyi}.

\bigskip
\begin{acknowledgments}
  
  MO and MCL acknowledge support from the International Human Frontier
  Science Program Organization (Grant RGY0069/2009-C).  The work of
  MRF, MCL, FH, and PT was supported by a ``Convergence'' grant from
  the University Pierre and Marie Curie, Paris.
  
 	We thank G. Malaguti, L. Peliti and S. Wielgoss
 	for the useful comments and discussions on this work.
\end{acknowledgments}

\setcounter{figure}{0}

\appendix
\numberwithin{figure}{section}

\section{Definition of the three model variants}
\label{appendix:modelli}

This Appendix describes in additional detail the three model variants
used in the main text.

All variants start from the main assumption that the selection
coefficient is dependent on the number of mutations $s=s_0 g'(k)$,
where $g'(k)$ is a decreasing function of $k$, and corresponds to
different specifications of $g'(k)$.  Every individual with $k$
beneficial mutation has fitness
\begin{equation}
  w_k=e^{ \sum_{k'=0}^k s_0 g'(k') } = e^{s_0 g(k)},
\label{fitness_def_App}
\end{equation}
and we can always arbitrarily define $w(0)=1$ as the ``wild-type'' fitness.
The main example of $g'(k)$ we considered is given by the choice of a
fitness gain that depends on the number of mutations occurred $k$ as a
power law.  In this case, $g'(k) = \alpha k^{\alpha -1}$ with $\alpha \le 1$,
where the epistasis grows in strength decreasing $\alpha$ from the
non-epistatic case of $\alpha=1$. In this case, the fitness is
\begin{equation}
  w_k=e^{ \sum_{k'=0}^k s_0 \alpha k'^{\alpha-1}} \ ,
\end{equation}
with $\alpha<1$ for diminishing return, and $\alpha=1\rightarrow$ for
no epistasis. 

In the general case $\alpha \ne 0$, one can write
 \begin{equation}
   w_k=e^{ s_0 \alpha H_{k,1-\alpha}} \approx 
   e^{ s_0 \left( k^{\alpha-1}\left[ \frac{k}{\alpha} + \frac{1}{2} +
         O(1/k) \right]   +\zeta(1-\alpha) \right) } \ ,
\end{equation}
where $\zeta$ indicates the Riemann zeta function. Thus, the relative
fitness can be expressed as
\begin{equation}
  \chi_k \approx e^{ s_0 \left(k^\alpha - \langle k^\alpha \rangle \right)}
\end{equation}

For the particular case $\alpha=0$,
\begin{equation}
  w_k=e^{ \sum_{k'=1}^k s_0 k'^{-1}} \approx  e^{s_0\left(ln(k) +\gamma
    \right)} \ ,
\end{equation}
which is obtained truncating the harmonic number expansion, and where
$\gamma=0.57721$ is the Euler-Mascheroni constant. This approximation
neglects the terms $O(1/k)$. Under this assumption, the relative
fitness can be expressed as
\begin{equation}
  \chi_k \approx e^{s_0 ln\frac{k}{\langle k\rangle}}
\end{equation}

These expressions show that it is essentially equivalent to directly
assume fitness functions of the form
\begin{eqnarray}
w_k=e^{s_0 k^\alpha} ~~\textrm{if } \alpha \ne 0\nonumber\\
w_k=e^{s_0 ln(k+1)} ~~\textrm{if } \alpha = 0  \ ,
\label{Fitness_Pow_Log_appendix}
\end{eqnarray}
that can be expressed, neglecting terms of order $1/k$, as sum of
powers of the number of accumulated mutations.  In the case $\alpha
=0$ the factor $\gamma$ results a multiplicative factor that can be
canceled since it does not affect the relative fitness, that is the
relevant quantity for the dynamics.  On the other hand, this notation
requires an appropriate rescaling of $s_0 \alpha \rightarrow \alpha$
in the case of $\alpha \ne 0$.  Under these assumptions, the relative
fitness, which ultimately defines the dynamics, is identical to the
one obtained considering the explicit sum of the contributions of
single mutations, neglecting corrections for small $k$.  Moreover, the
fitness functions~\ref{Fitness_Pow_Log_appendix}, using $ln(k+1)$
instead of $ln(k)$, allow to automatically include the case $k=0$ with
the correct normalization condition $w(0)=1$.
We refer to the case $\alpha \ne 0$ as the power law model,
while the case $\alpha=0$ corresponds to the logarithmic model. 	

In the case of an approximately Gaussian distribution of mutation
classes with distance $L_k$ between the fittest class and the class
with mean log-fitness, the relative fitness of the fittest class can
be expressed as
\begin{equation}
\chi_k = e^{  s_0(k^\alpha - ( k -L_k)^\alpha )} 
\approx 
e^{s_0 	\alpha L_k k^{\alpha-1}} \ ,
\end{equation}
where the expansion is performed for $L_k/k \ll 1$ which should be the
case in our system. In the main text, we use this approximation to
perform the estimate of the adaptation speed.  In the non-epistatic
case ($\alpha=1$) the expression gives the relative advantage
expression of the fittest class used in the standard multiple mutation
model. Similarly, in the limit $\alpha=0$
 \begin{equation}  
\chi_k \approx e^{s_0 L_k /k} \ .
\end{equation}

The third model we use is characterized by a ``geometric dependence'' of
the fitness advantage on the number of acquired mutations. In this
case the advantage function is $g'(k)=q^{k-1}$ with $q<1$ and the
fitness is given by 
\begin{equation}
  w_k=e^{ \sum_{k'=1}^k s_0 q^{k'-1}} = e^{s_0 \frac{1-q^k}{1-q}} .
\end{equation}
In other words, the advantage accumulates following a geometric sum.
As for the former models, considering only the final form of the
advantage function $g(k)=\frac{1-q^k}{1-q}$, allows to directly
satisfy the condition $g(0)=0$.  The constant factor $1-q$ could be
adsorbed in $s_0$, as for the power law model.  In the same
quasi-Gaussian approximation described for the power law and
logarithmic model, the relative fitness is given by
\begin{equation}  
  \chi_k \approx e^{s_0 \frac{q^{\langle k \rangle} -q^k}{1-q}},
\end{equation}
and, in the case of the fittest class it becomes
\begin{equation}  
  \chi_k \approx e^{s_0 q^{\langle k \rangle} \frac{1 -q^L}{1-q}} \ .
\end{equation}

Note that, while in the previous models the fitness is not bounded
from above, the fitness for the geometric model is finite for infinite
$\langle k \rangle$ (i.e. $w(k\rightarrow
\infty)=e^{s_0\dfrac{1}{1-q}}$).

	\begin{figure}[htbp]
	\centering
    \includegraphics[width=0.45\textwidth]{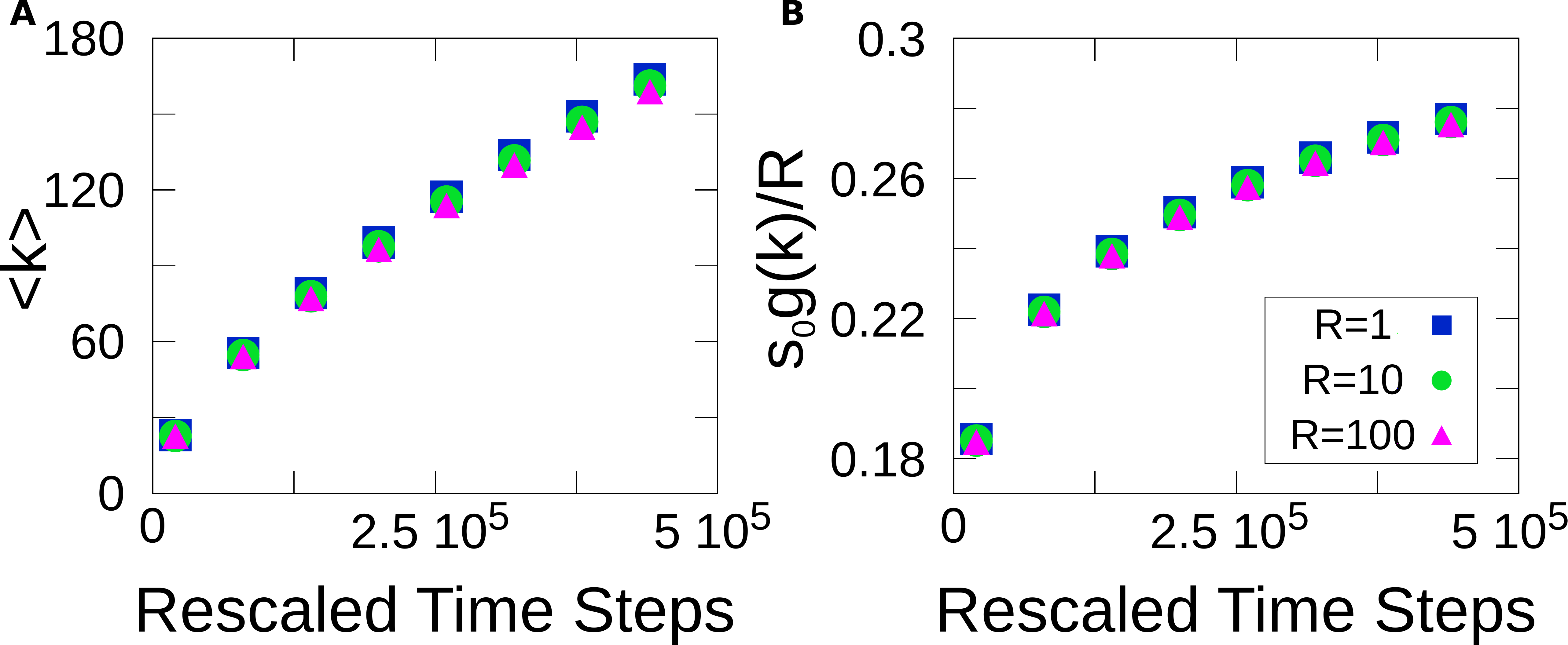} 
    \caption{(Color online)\textbf{The model is effectively invariant for rescaling
        of the parameters. } The figure shows the mean number of
      mutations ($\langle k \rangle$, panel A) and the fitness
      advantage (panel B) obtained from simulations run using three
      different rescaling factors ($r$=1,10,100 different symbols as
      in legend).  The simulated dynamics is almost unaffected by the
      rescaling procedure. Simulations are made using the parameters
      $N=10^9$, $s_0=0.1$, $\alpha=0.2$, $U_b=2\cdot10^{-6}$. The data
      are averaged over $100$ iterations, and error bars (standard
      error) are smaller than symbols.}
  \label{fig_R}
\end{figure}

	\begin{figure}[htbp]
	\centering
    \includegraphics[width=0.40\textwidth]{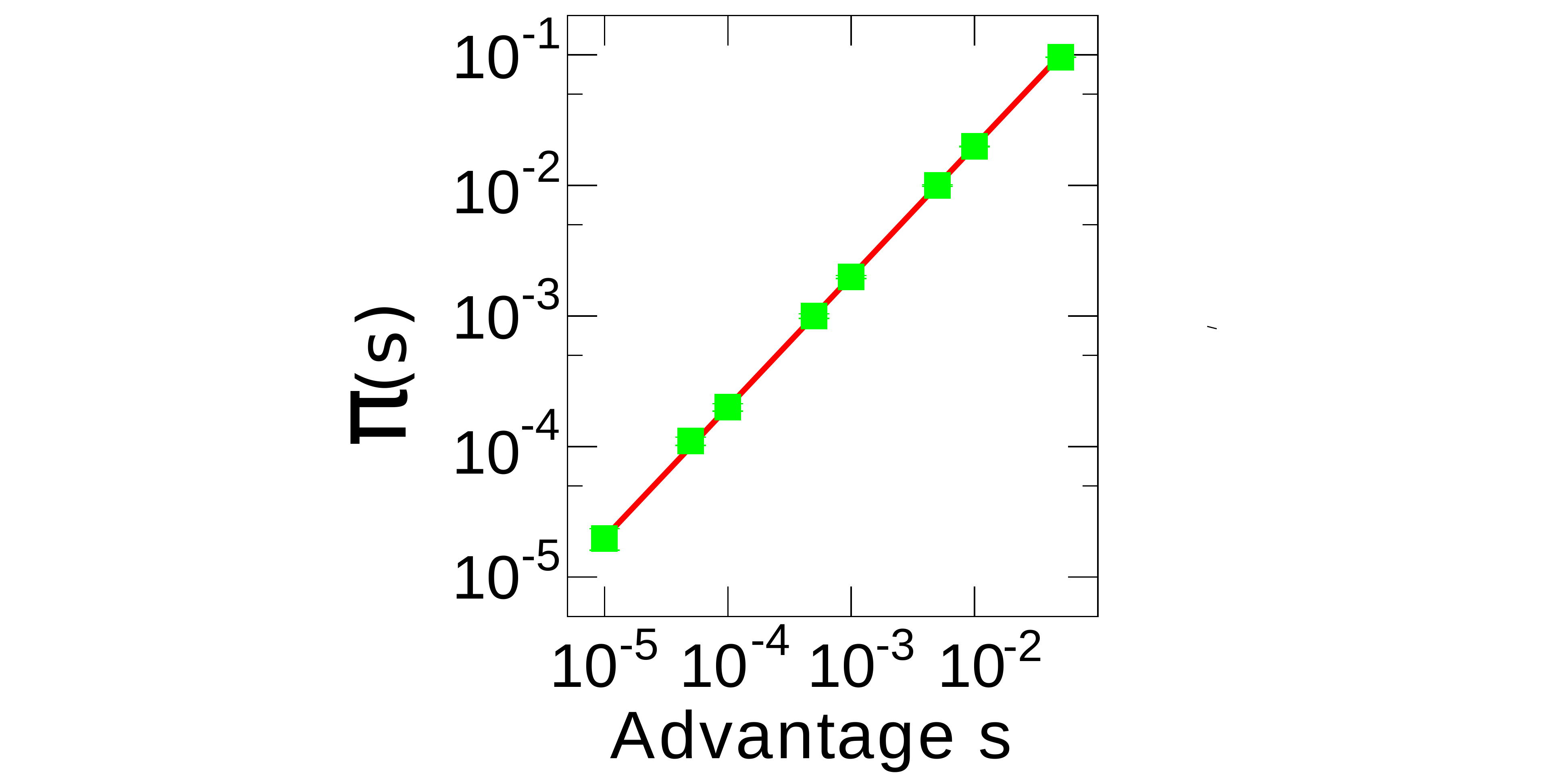} 
    \caption{\textbf{The fixation probability is proportional to the
        fitness advantage.} The figure shows the fixation
      probability $\pi(s)$ of a single clone that grows in a uniform
      background having advantage $s$ (symbols), measured from our
      simulations.  The fixation probability is obtained as the ratio
      between the number of realizations where the beneficial mutator fixes
      and the total number of realizations ($10^8$). The
      results give $\pi(s)=2s$ (continuous red line), in accordance
      with~\cite{ParkKrug}. Simulations are performed using $N=10^7$.
    }
  \label{fig_PI}
\end{figure}

	\begin{figure}[htbp]
	\centering
    \includegraphics[width=0.45\textwidth]{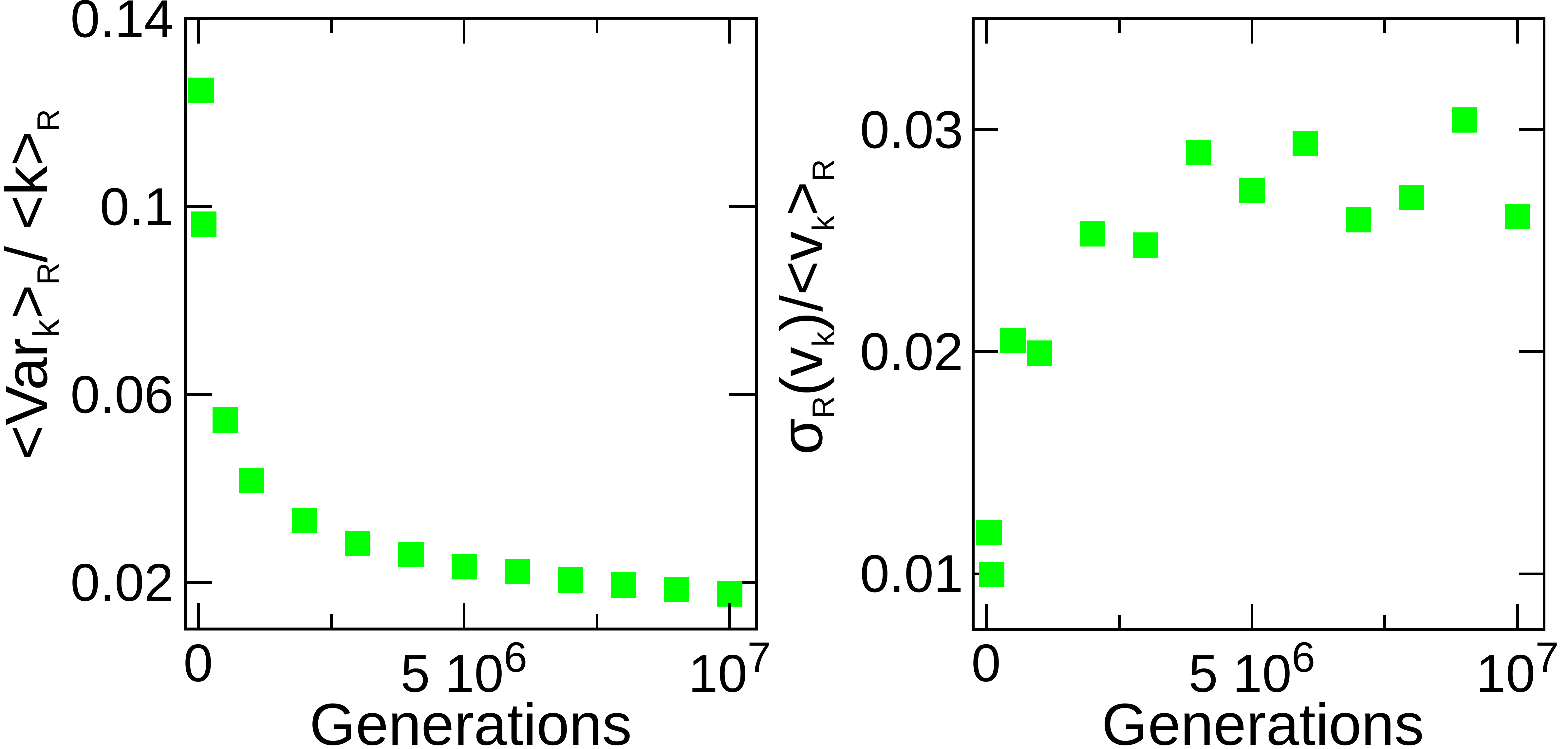} 
    \caption{\textbf{The relative variance of the distribution
        decreases in time, while the relative variance of the velocity
        follows an increasing trend. } The figure shows the ratio
      between the variance $\mathrm{Var}_k$ and the mean number of
      mutations $\langle k \rangle$ (left panel) and the ratio between
      $\mathrm{Var}_{v_k}$ (calculated over different realizations)
      and the velocity $v_k$ (right panel) as a function of time.
      Simulations results confirm that, despite $\mathrm{Var}_k$
      increases in time, the mean number of mutations increases more
      rapidly. Since $D_k \leq L_k$,
      simulations results confirm the hypothesis
      $D_k/\langle k\rangle \ll1$ used in the main text (see also
      Fig.~\ref{Fig_Gaussiane}B).  The increase 
      in variance of the velocity reflects both the increase of the
      variance of the distribution and the decrease of $v_k$. However
      the error is  sufficiently small not to affect the results on 
      experimentally relevant time scales.
      Simulations are performed using the parameters $N=10^7$,
      $s_0=0.5$, $\alpha=0.02$, $U_b=1\cdot10^{-3}$. Data are averaged
      over $100$ realizations of the process.}
  \label{fig_Var}
\end{figure}

\section{Experimental data and definition of fitness}
\label{appendix:experiment}

This section briefly summarizes the main features of the evolutionary
experiments with \emph{A.~baylyi} and \emph{E.~coli} considered here,
in order to describe the assumptions made to link genomic and growth
rate measurements to fitness in the model.
More detailed information about the long-term evolution experiment
with \emph{E.~coli}, and in particular about fitness measurements can
be found in refs.~\cite{Lenski1991, LTEE_Ecoli}.  Fitness data used
during the parameter-matching procedure are reported in
ref.~\cite{deVisser2002, Barrick2009}.  A comprehensive description of
the experimental methods used for the experiments with \emph{A.~baylyi}
can be found in ref.~\cite{Thomen2012}.
Note that in our modeling framework, as in most standard evolutionary
models, the population size is kept constant and all individuals are
substituted by newly generated offsprings at every generation,
assuming a constant time interval between generations, although none
of these assumptions are completely verified in the two experiments we
considered.

\subsection{Acinetobacter baylyi evolution experiment}

In the \emph{A.~baylyi} experiment the population size is almost
constant and, even if generations are not synchronous, the mean growth
rate is held fixed by the dilution rate in the chemostat.  Maximum
growth-rate measurements were performed in batch on single-clone
colonies.  These values differ from the effective growth rate that
each population can reach in chemostat because of competition between
different sub-populations for a limited amount of nutrients.  However,
the measured values of $\mu_{max}$ are supposed to be indicative of
the effective fitness of the different sub-populations inside the
chemostat.

When a monoclonal population grows in the chemostat, its growth rate
equals the dilution rate $D$, while in presence of different
sub-populations the dynamics becomes more
complex~\cite{Dykhuizen1983}.  For the sake of simplicity, and since
the detailed dynamics is not experimentally accessible, we assume that
the growth rate of all clones is fixed by dilution rate in the
chemostat over the whole experiment and that the generation time is
defined as $t_{gen}= ln(2)/D$.  Thus, the different maximum growth
rates of clones measured in batch are interpreted as different
survival probabilities of their offspring inside the chemostat.

Defining the fitness as the mean expected number of (surviving)
offspring per generation inside the chemostat we assume that it is
estimated by the maximum growth rate of mutant, i.e.  $w(i)=
e^{m_i}=e^{\mu_{max,i} t_{gen}}$, where the index $i$ indicates the
$i-th$ sub-population, $m_i$ is the growth rate expressed in
$1/\mathrm{generation}$, experimental data of $\mu_{max,i}$ have the
units of $h^{-1}$, and $t_{gen}= ln(2)/D \approx 1h^{-1}$
($D\approx0.7h^{-1}$). In simple words the fitness of an individual is
defined as the amount of offspring assuming it can grow by the maximum
growth rate over an average generation defined by the chemostat
dilution (and thus fitness is a dimensionless quantity).
Since the relevant quantity of the dynamics is the relative fitness,
during the parameter-matching procedure we used the normalized fitness
($w(i)=e^{\mu_{max,i} - \mu_{max, WT}}$) to infer the functional form
of the advantage.

\subsection{Escherichia coli evolution experiment}

The long-term \emph{E.~coli} evolution experiment is performed in
batch, and serial dilution $1:100$ is performed daily. The maximum
population size ($\approx 5\cdot10^8$) is fixed by the total amount of
nutrient in the medium. The number of doubling for each day is
$log_2(100)\approx 6.64$ as derived from the 100-fold daily
increase~\cite{Lenski1991}.  
The effective population size can be estimated around $2\cdot10^7$
individuals. Since the speed of evolution depends logarithmically on
the population size, the error on the estimate of $U_b$ given by this
approximation should be very low.
Competition experiments are performed between samples at different
times and a spontaneous mutant of the ancestor, which is easier to
track visually and has been verified to have almost the same
fitness~\cite{deVisser2002}.
Relative log-fitness of population $i$ respect to the wild type is defined as
\begin{equation}
\varphi(i)= \dfrac{\ln\left(\dfrac{f(1)_i}{f(0)_i}\right)}
{\ln\left(\dfrac{f(1)_{WT}}{f(0)_{WT}}\right)},
\end{equation} 
where $f(1)$ and $f(0)$ indicate the frequencies at the beginning and
the end of the competition experiment~\cite{deVisser2002,Barrick2009}.
 $\varphi(i)$ can be seen as
the ratio between the growth rate of the two populations and
corresponds to log-fitness in the model. Note again that the
experimental data are dimensionless.
The difference between the Malthusian parameters can be
approximatively deduced as $(\varphi(i)-1)\bar{\mu} \approx (\mu_i -
\mu_{WT})$ where $\bar{\mu}$ is the mean Malthusian
parameter~\cite{Lenski1991}.
Since we are interested in expressing the fitness as the mean number
of offspring per generation, we use as mean Malthusian parameter
$\bar{\mu}=\ln(2) \mathrm{generations}^{-1}$.  Thus we define the normalized
fitness as $w(i)=e^{(\varphi(i)-1)\ln(2)}$.

\bibliography{DiminishingRetInterf.bib}

\end{document}